\documentclass{ptephy}

\newcommand{\ngt}{\lower.5ex\hbox{$\; \buildrel > \over \sim \;$}}
\newcommand{\nlt}{\lower.5ex\hbox{$\; \buildrel < \over \sim \;$}}

\begin{document}

\title{Cosmological and Astrophysical Implications of the
Sunyaev-Zel'dovich Effect}

\author{\name{Tetsu Kitayama}{1}}

\address{\affil{1}{Department of Physics, Toho University,  
Funabashi, Chiba 274-8510, Japan}}

\begin{abstract}
The Sunyaev-Zel'dovich effect provides a useful probe of cosmology and
structure formation in the Universe. Recent years have seen rapid
progress in both quality and quantity of its measurements.  In this
review, we overview cosmological and astrophysical implications of
recent and near future observations of the effect. They include
measuring the evolution of the cosmic microwave background radiation
temperature, the distance-redshift relation out to high redshifts,
number counts and power spectra of galaxy clusters, distributions and
dynamics of intracluster plasma, and large-scale motions of the
Universe.
\end{abstract}

\subjectindex{cosmology, clusters of galaxies, radio observations}

\maketitle

\section{Introduction}

The Sunyaev-Zel'dovich effect (SZE, \cite{zs69,sz70,sz72,sz80a}) is inverse
Compton scattering of the Cosmic Microwave Background (CMB) photons off
electrons in clusters of galaxies or any cosmic structures. It is
amongst major sources of secondary anisotropies of the CMB on sub-degree
angular scales. The most noticeable feature of the SZE is that its
brightness is apparently independent of the source redshift $z$ because
its intrinsic intensity increases with redshifts together with the
energy density of seed (CMB) photons; otherwise observed brightness
should decrease rapidly as $(1+z)^{-4}$. This makes the SZE a unique
probe of the distant Universe. The SZE also has a characteristic
spectral shape which helps separating it from other signals such as
radio galaxies and primary CMB anisotropies.

\begin{figure}[th]
\centering\includegraphics[width=85mm]{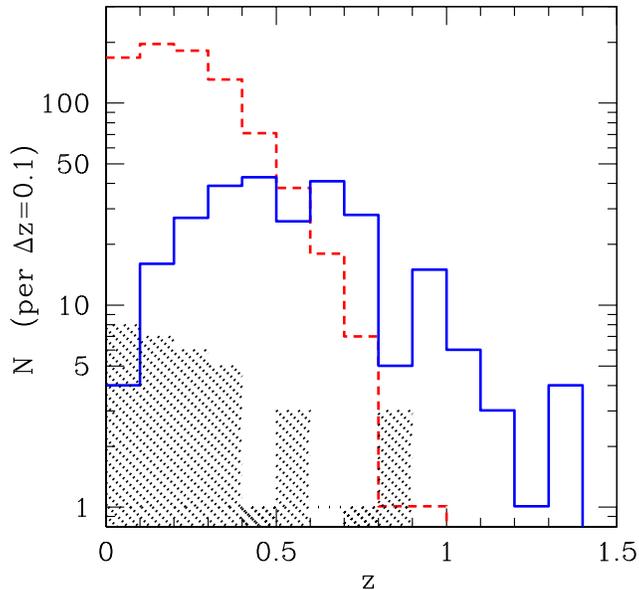} \caption{
 Redshift histograms of galaxy clusters with measured thermal SZE
 signals and redshifts from the literature. Solid line indicates 258
 clusters detected in the ground-based surveys either by SPT
 \cite{spt09,spt10,spt11,spt13a} or by ACT \cite{act10,act11,act13a},
 excluding overlaps, over a total of $\sim 3000$ deg$^2$. Dashed line
 shows 813 clusters detected by the Planck satellite over all sky
 \cite{planck_earlysz,planck13a}. For reference, hatched region marks 34
 clusters with $>4\sigma$ SZE detections published as of 2002
 (\cite{birkinshaw99,carlstrom02} and references therein) which consist
 mainly of X-ray luminous clusters.}  \label{fig-nsz}
\end{figure}

Recent developments of large area surveys by the South Pole Telescope
(SPT) \cite{spt09,spt10,spt11,spt13a}, the Atacama Cosmology Telescope
(ACT) \cite{act10,act11,act13a}, and the Planck satellite
\cite{planck_earlysz,planck13a} have enlarged the sample of galaxy
clusters observed through the SZE by more than an order of magnitude
over the last decade as illustrated in Figure \ref{fig-nsz}. To
date, the SZE by thermal electrons (thermal SZE) has been detected for
about $1000$ galaxy clusters including more than 200 new clusters
previously unknown by any other observational means. The imprint of yet
unresolved smaller-scale cosmic structures has been explored by
means of their angular power spectrum
\cite{reichardt12,sievers13,planck_y} and the stacking analysis
\cite{hand11,planck_group}.  There have been reports of detections of
the SZE by peculiar motions of galaxy clusters (kinematic SZE) either
statistically \cite{hand12} or from a high-velocity merger
\cite{sayers13b}.

The sensitivity of the SZE observations of individual clusters has
also improved significantly, making it a useful tool for studying
physics of intracluster plasma. In particular, the SZE data provide a
direct measure of thermal pressure of electrons, which is highly
complementary to X-ray observations.  They allow us to study the
distance-redshift relation (e.g., \cite{schmidt04,bonamente06}),
three-dimensional structures \cite{defilippis05,sereno12}, and dynamics
\cite{kitayama04,korngut11,planck_coma} of galaxy clusters. By means of
the SZE, we are witnessing the high-mass end of structure formation in
the Universe that in turn serves as a powerful probe of cosmology.

Theoretical foundations and earlier observations of the SZE are reviewed
extensively by \cite{sz80b,rephaeli95,birkinshaw99,carlstrom02}.  In the
present paper, we focus mainly on practical applications of the SZE
that have become more feasible by recent observations, and discuss their
cosmological and astrophysical implications. Unless explicitly stated
otherwise and wherever necessary to assume specific values of
cosmological parameters, we adopt a conventional $\Lambda$CDM model
with the matter density parameter $\Omega_{\rm m}=0.3$, the dark energy
density parameter $\Omega_{\Lambda}=0.7$, the baryon density parameter
$\Omega_{\rm b}=0.045$, the Hubble constant $h_{70}=H_0/
(70 \mbox{\:km\:s$^{-1}$Mpc$^{-1}$})=1.0$, the dark energy equation of state parameter
$w=-1.0$, the amplitude of density fluctuations $\sigma_8=0.8$, and the
spectral index of primordial density fluctuations $n_{\rm s}=0.96$.

\section{The Sunyaev-Zel'dovich Effect} 
\label{sec-sz}

When CMB photons pass through a cloud of free electrons with number
density $n_{\rm e}$, they are subject to scattering with a probability
characterized by the optical depth
\begin{equation}
\tau_{\rm e} = \int \sigma_{\rm T} n_{\rm e} dl  \,\sim \, 2 \times 10^{-3} 
\left(\frac{n_{\rm e}}{10^{-3} \mbox{ cm}^{-3}}\right)
\left(\frac{l}{\mbox{Mpc}}\right), 
\label{eq-taue}
\end{equation}
where $\sigma_{\rm T}$ is the Thomson cross section, $\int \cdots dl$
denotes the line-of-sight integral, and the quoted values are typical of
galaxy clusters. It follows that a single scattering is in general a
good approximation even in largest galaxy clusters. The Thomson limit
applies in the rest frame of an electron as long as its velocity $v_{\rm
e}$ relative to the CMB satisfies $\gamma_{\rm e} \equiv (1-v_{\rm
e}^2/c^2)^{-1/2} \ll m_{\rm e}c^2/(k_{\rm B}T_{\rm CMB}) \sim
10^9/(1+z)$, where $m_{\rm e}$ is the electron mass, $c$ is the speed of
light, $k_{\rm B}$ is the Boltzmann constant, and $T_{\rm CMB}$ is the
CMB temperature. In the CMB rest frame, on the other hand, the net
energy is transferred from the electron to the photons for $v_{\rm e}
\gg \sqrt{k_{\rm B}T_{\rm CMB}/m_{\rm e}}\sim 10 (1+z)^{1/2}$ km
s$^{-1}$ and energies of the scattered photons increase by a factor of
$\sim \gamma_{\rm e}^2$ on average. While such {\it inverse} Compton
scattering takes place in a wide range of cosmic plasma, the term SZE is
conventionally used for scattering of the CMB photons at GHz to THz
frequencies
by non-relativistic or mildly relativistic
electrons.  Its intensity is often expressed by a series of $v_{\rm
e}/c$.

To the lowest order in $v_{\rm e}/c$, an apparent small change in
$T_{\rm CMB}$ by the above scattering corresponds to the Doppler effect
\cite{sz72,sz80a} and called the kinematic SZE:
\begin{eqnarray}
\frac{\Delta T_{\rm CMB}}{T_{\rm CMB}}
= \int \sigma_{\rm T} n_{\rm e} \frac{v_{\parallel}}{c} 
dl \, \sim \, 7 \times 10^{-6} 
\left(\frac{n_{\rm e}}{10^{-3} \mbox{ cm}^{-3}}\right)
\left(\frac{v_{\parallel}}{10^3 ~\mbox{km\:s$^{-1}$}}\right)
\left(\frac{l}{\mbox{Mpc}}\right),
\label{eq-ksz2}
\end{eqnarray}
where $v_{\parallel}$ is the line-of-sight component of $v_{\rm e}$ and
is taken to have a positive value toward the observer. Equivalently, the
CMB intensity spectrum is distorted by
\begin{eqnarray}
\frac{\Delta I_\nu}{I_{\rm \nu,CMB}} 
= \frac{x e^{x}}{e^{x}-1} \int \sigma_{\rm T} n_{\rm e}
\frac{v_{\parallel}}{c} dl,   
\label{eq-ksz} 
\end{eqnarray}
where $I_{\rm \nu,CMB} = i_0 x^3/(e^{x}-1)$, $i_0 = 2(k_{\rm B} T_{\rm
CMB})^3/(h_{\rm P} c)^2$,
$h_{\rm P}$ is the Planck constant, and $x$ is the dimensionless 
photon frequency $\nu$ defined by  
\begin{eqnarray}
x &=& \frac{h_{\rm P} \nu}{k_{\rm B} T_{\rm CMB}}
= 1.76 
\left(\frac{\nu}{100 \mbox{~GHz}}\right)
\left(\frac{T_{\rm CMB}}{2.726 \mbox{~K}}\right)^{-1}.
\label{eq-xcmb}
\end{eqnarray}
Note that random velocities cancel out in equations (\ref{eq-ksz2}) and
(\ref{eq-ksz}) and the coherent motion with respect to the CMB is
responsible for the kinematic SZE.

\begin{figure}[th]
\centering\includegraphics[width=85mm]{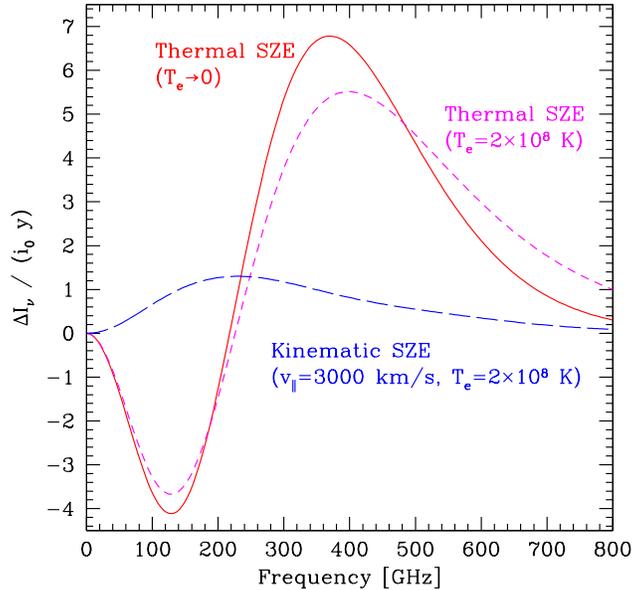}
 \caption{Spectra of the Sunyaev-Zel'dovich effect (SZE). Intensity
 differences from the CMB normalized by $i_0 y$ are plotted for the
 non-relativistic thermal SZE (solid), the thermal SZE with the
 relativistic correction \cite{itoh04} for $T_{\rm e}=2\times 10^8$ K
 (short dashed), and the kinematic SZE with the relativistic correction
 \cite{nozawa06} for the bulk velocity $3000$ km\:s$^{-1}$ toward us and
 $T_{\rm e}=2\times 10^8$ K (long dashed). The ratio between the
 kinematic SZE and the thermal SZE is proportional to
 $v_{\parallel}/T_{\rm e}$ in the non-relativistic limit.}
 \label{fig-sz}
\end{figure}

Isotropic random motions of Maxwellian electrons, on the other hand,
give rise to the thermal SZE \cite{zs69,sz70, sz72}. For electrons with
temperature $T_{\rm e}$ ($\gg T_{\rm CMB}$), the leading term of the
spectral distortion is of order $(v_{\rm e}/c)^2 \propto k_{\rm B}
T_{\rm e}/(m_{\rm e}c^2)$ and given by
\begin{eqnarray}
\frac{\Delta T}{T_{\rm CMB}}&=& \left( x \coth \frac{x}{2} -4
				       \right) \, y, 
\label{eq-tsz2} 
\end{eqnarray}
or equivalently,  
\begin{eqnarray}
\frac{\Delta I_\nu}{I_{\rm \nu, CMB}} 
&=& \frac{x e^{x}}{e^{x}-1} \left( x \coth \frac{x}{2} -4
				       \right)   \,y, 
\label{eq-tsz} 
\end{eqnarray}
where $y$ is the Compton y-parameter:  
\begin{eqnarray}
\label{eq-y}
y = \int  
\sigma_{\rm T} n_{\rm e} \frac{k_{\rm B} T_{\rm e}}{m_{\rm e} c^2} dl 
\, \sim  \, 4 \times 10^{-5} 
\left(\frac{n_{\rm e}}{10^{-3} \mbox{ cm}^{-3}}\right)
\left(\frac{T_{\rm e}}{10^8 \mbox{ K}}\right)
\left(\frac{l}{\mbox{Mpc}}\right), 
\end{eqnarray}
which is essentially the dimensionless integrated electron pressure
$P_{\rm e}=k_{\rm B} n_{\rm e} T_{\rm e}$.  

Although the thermal SZE is of second order in $v_{\rm e}/c$, it
dominates over the kinematic SZE typically by an order of magnitude for
galaxy clusters because the thermal velocity of electrons $\sqrt{k_{\rm
B}T_{\rm e}/m_{\rm e}} \sim 4 \times 10^4 (T_{\rm e}/10^8
\mbox{K})^{1/2}$ km\:s$^{-1}$ is much larger than bulk velocities of
$\nlt 10^3$ km\:s$^{-1}$. In the non-relativistic limit, spectral
shapes of the kinematic SZE and the thermal SZE (eqs [\ref{eq-ksz}] and
[\ref{eq-tsz}]) depend only on $x$. Corrections due to higher order
terms are non-negligible once electrons become relativistic
\cite{wright79,fabbri81,rephaeli95a,challinor98,itoh98,sazonov98,nozawa98}.
The spectral shape of the thermal SZE then starts to depend on $T_{\rm
e}$ and that of the kinematic SZE on both $T_{\rm e}$ and the bulk
velocity. In any case, observed amplitude and spectral shape of the SZE
are both independent of $z$, because $\Delta T$ and $\Delta I_\nu$ are
redshifted in exactly the same way as $T_{\rm CMB}$ and $I_{\rm \nu,
CMB}$, respectively.

\begin{figure}[t]
\centering\includegraphics[width=153mm]{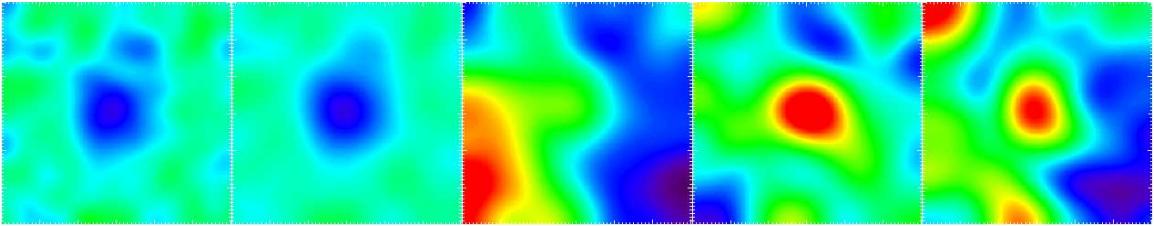}
 \caption{Cleaned images of Abell 2256 at $z=0.058$ observed by Planck at
 100, 143, 217, 353, and 545 GHz from left to right
over a size of $1^\circ \times 1^\circ $ (reproduced from
\cite{planck_earlysz} with permission, \copyright\,ESO); $1^\circ = 4.0 
\,h_{70}^{-1}$Mpc in the cluster rest frame.  Blue, green, and red
colors indicate negative, null, and positive intensities with respect to
the CMB, respectively. The FWHMs of observing beams are $9.5'$, $7.1'$,
$4.7'$, $4.5'$, and $4.7'$ from left to right \cite{planck_hfi}.}
\label{fig-szimage}
\end{figure}

Figure \ref{fig-sz} illustrates spectral shapes of the thermal SZE and
the kinematic SZE for representative values of $T_{\rm e}$ and
$v_{\parallel}$.  The thermal SZE leads to a decrement at $\nu<218$ GHz
($x<3.83$) and an increment at higher frequencies. The relativistic
correction shifts the null of the thermal SZE and modifies the
spectral shape especially at high frequencies. The kinematic SZE, on the
other hand, has its peak near the null of the thermal SZE.
Multi-frequency measurements are necessary to separate the kinematic SZE
from the thermal SZE and/or to determine $T_{\rm e}$ via the
relativistic correction.  Figure \ref{fig-szimage} further shows real
images of a galaxy cluster, Abell 2256, taken by Planck
\cite{planck_earlysz}. Both decrement and increment signals of the
thermal SZE are detected clearly at low and high frequencies,
respectively.  The data are also consistent with the null of the thermal
SZE at 217 GHz, with no apparent signature of the kinematic SZE.

Small amounts of polarization are produced by inverse Compton scattering
owing to anisotropies of the radiation field in the electron rest frame
\cite{sz80a,audit99,sazonov99,challinor00,itoh00}. Leading effects are
due to i) the CMB quadrupole with the maximum polarization degree of
$\sim 10^{-6} \tau_e$ toward the sky directions that are perpendicular
to the quadrupole plane, ii) the transverse velocity $v_\perp$ of
electrons on the sky with the polarization degrees of $\sim 0.1
(v_\perp/c)^2 \tau_{\rm e}$ and $\sim 0.01 (v_\perp/c) \tau_{\rm e}^2$,
and iii) thermal electrons with the polarization degree of $\sim 0.01
(k_{\rm B}T_{\rm e}/m_{\rm e}c^2) \tau_{\rm e}^2$; the prefactors are in
the Rayleigh-Jeans limit and their frequency dependence as well as
angular distribution can be found in \cite{sazonov99}. The effects
proportional to $\tau_{\rm e}^2$ originate from multiple scatterings and
are more sensitive to the spatial distribution of electrons. While all
the effects are beyond the sensitivity of current detectors, they
contain unique cosmological and astrophysical information.  The first
effect will provide a knowledge of the CMB quadrupole as seen by
clusters including those at high redshifts, thereby reducing the cosmic
variance uncertainty \cite{kamionkowski97}. The second effect, together
with the kinematic SZE, will in principle offer a measure of the 3D
velocity of the gas. The third effect will allow us to separate $T_{\rm
e}$ and $\tau_{\rm e}$ in the thermal SZE.

Nonthermal electrons also upscatter the CMB photons. A number of galaxy
clusters host diffuse synchrotron emission from relativistic electrons
with $\gamma_{\rm e}\sim 10^4$ (see \cite{feretti12} for a review);
while inverse Compton scattering by the same population of electrons
should emerge in hard X-rays, their lower energy counterparts, if 
present, give rise to the nonthermal SZE.  Predicted spectral
distortions of the CMB, particularly at high frequencies, are sensitive
to the underlying energy distribution of nonthermal electrons
\cite{birkinshaw99,ensslin00,blasi00,colafrancesco03}.  Major
difficulties in actually observing them are the short life-time of
suprathermal electrons as well as a large amount of contamination
including the thermal SZE itself and dusty galaxies.

\section{Evolution of the CMB Temperature}

A fundamental prediction of the standard cosmology is that the CMB
temperature evolves with redshift adiabatically as
\begin{eqnarray}
 T_{\rm CMB}(z)=
T_{\rm CMB}(0) ~ (1+z), 
\label{eq-tcmbz}
\end{eqnarray}
where $T_{\rm CMB}(0)=2.7260\pm 0.0013$ K is the present-day CMB
temperature measured by COBE-FIRAS \cite{fixsen09}.  Any deviation from
the above evolution would be a signature of a violation of conventional
assumptions such as the local position invariance and the photon number
conservation.

In fact, redshift-independence of the spectral shape of the SZE
mentioned in Section \ref{sec-sz} also relies on equation
(\ref{eq-tcmbz}). This in turn makes it possible to use the observed SZE
spectra to measure the CMB temperature
\cite{fabbri78,rephaeli80,lamagna07} at an arbitrary $z$ and test the
validity of equation (\ref{eq-tcmbz}). A great advantage of this method
is that it uses only the spectral shape of the thermal SZE and does not
rely, at least in principle, on details of underlying gas properties. In
practice, appropriate gas distributions should be taken into account to
correct for beam dilution effects at different observing frequencies.
As long as the large-scale bulk motion is small (see Section
\ref{sec-unres}), the kinematic SZE would primarily increase the
dispersion of the measurement. The relativistic effects on the SZE can
be suppressed by using clusters with relatively low electron
temperatures; accurate temperature measurements of individual clusters
will be necessary otherwise.  Multi-frequency data will also be crucial
for separating contamination by radio sources, CMB primary anisotropies,
and dust emission.

\begin{figure}[tbp]
\centering\includegraphics[width=85mm]{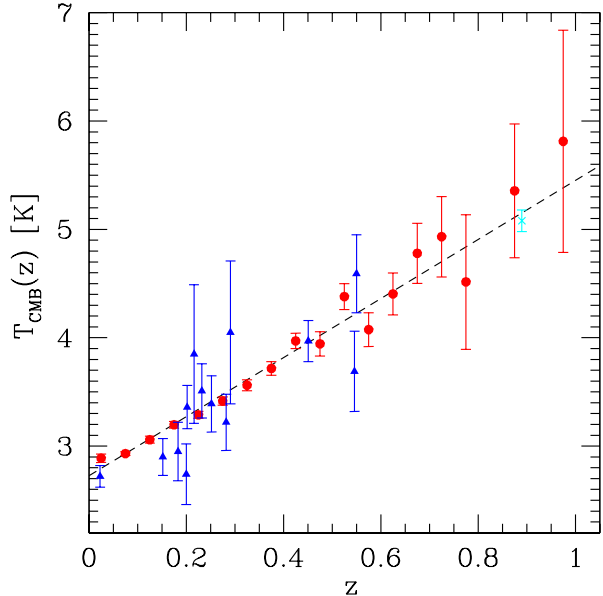} \caption{Measured
redshift evolution of the CMB temperature as compared to the adiabatic
prediction (dashed line).  Red circles indicate the SZE measurements by
stacking a sample of clusters detected by Planck \cite{hurier13}; note
that there are fewer clusters in each redshift bin at higher $z$ (see
Fig. \ref{fig-nsz}).  Blue triangles show the SZE measurements for
individual clusters \cite{luzzi09}. A cyan cross is the measurement
using molecular absorption lines \cite{muller13}.}  \label{fig-tz}
\end{figure}

A conventional approach of generalizing
equation (\ref{eq-tcmbz}) is to adopt the form
$T_{\rm CMB}(z)=
T_{\rm CMB}(0) ~ (1+z)^{1-\beta_{\rm T}}$
\cite{lima00} and to determine the parameter $\beta_{\rm T}$ from the
data. Recent SZE measurements for a sample of galaxy clusters give
$\beta_{\rm T}=0.009 \pm 0.017$ at $z<1$ using the Planck data at
$100\sim 857$ GHz toward 813 clusters \cite{hurier13} and $\beta_{\rm
T}=0.017^{+0.030}_{-0.028}$ at $z < 1.35$ using the SPT data at 95 GHz
and 150 GHz toward 158 clusters \cite{saro13}. Figure \ref{fig-tz} shows
that current measurements are consistent with the adiabatic prediction,
while averaging over a large number of clusters is necessary owing to
the variance inherent to individual clusters.

Another independent measure of the CMB temperature at $z>0$ comes from
quasar absorption line spectra; if the relative population of the
different energy levels of atoms or molecules are in radiative
equilibrium with the CMB at that epoch, the excitation temperature of
the species gives a measure of $T_{\rm CMB}(z)$ \cite{bahcall68}. Major
sources of systematics are contributions of other heating sources, such
as collisions and local radiation fields. Combining transition lines of
various species are particularly useful for constraining the physical
conditions of the absorbing gas; e.g., a comprehensive analysis of a
molecular absorber at $z=0.89$ yields $T_{\rm CMB} = 5.08 \pm 0.10$ K
\cite{muller13} in agreement with the adiabatic expectation of 5.15 K
(Fig. \ref{fig-tz}).  At $1.7<z<2.7$, the CMB temperatures derived from
rotational excitation transitions of CO are also consistent with
equation (\ref{eq-tcmbz}) \cite{noterdaeme11}.

In summary, the high significance measurements currently available
are consistent with the adiabatic evolution of the CMB temperature and
we assume it throughout this paper.

\section{Distance Determinations}
\label{sec-dist}

It has long been recognized that the thermal SZE and the X-ray emission
from galaxy clusters provide a primary distance indicator that is
entirely independent of the cosmic distance ladder
\cite{cavaliere77,silk78,birkinshaw79,cavaliere79}.  This method employs
the fact that the SZE and the X-ray emission arise from the same thermal
gas but depend on its density in a different manner
(Sec. \ref{sec-xsz}).  Essentially the same method is readily applied to
testing the distance duality relation \cite{uzan04}.  Baryon fraction of
galaxy clusters measured by X-ray and/or SZE observations can also be
used to determine their distances \cite{sasaki96,pen97}
(Sec. \ref{sec-fgas}). Key assumptions in both methods are spherically
symmetric and smooth distribution of the gas and impacts of possible
violation of these assumptions are also discussed below.

\subsection{Combination with X-ray data}
\label{sec-xsz}

Suppose that a galaxy cluster at redshift $z$ has radial profiles of
electron density $n_{\rm e}(\phi)$ and temperature $T_{\rm e}(\phi)$
at the angular radius $\phi$ from its center, i.e., the physical
radius in three dimensional space divided by the angular diameter
distance $d_{\rm A}$ to the center. The X-ray surface brightness at the
projected angle $\theta$ on the sky from the center is given by the
line-of-sight integral:
\begin{eqnarray}
 I_{\rm X}(\theta) = \frac{2d^3_{\rm A}}{4 \pi d^2_{\rm L}}
  \int_{\theta }^{\infty} 
n^2_{\rm e}(\phi)\Lambda_{\rm X}[T_{\rm e}(\phi),Z(\phi),z]
  \frac{\phi d\phi}{\sqrt{\phi^2-\theta^2}}, 
\label{eq-ix1}
\end{eqnarray}
where $d_{\rm L}$ is the luminosity distance, $Z$ is the gas
metallicity, and $\Lambda_{\rm X}$ is the X-ray cooling function
including the k-correction, i.e., $n^2_{\rm e} \Lambda_{\rm X}$ stands
for the energy radiated per unit time and unit volume in the rest frame
of the cluster. In general, $\Lambda_{\rm X}$ depends only weakly on
$T_{\rm e}$ (weaker than $T_{\rm e}^{1/2}$) within the limited energy
band and a combination with X-ray spectral data allows one to measure
$T_{\rm e}(\phi)$, $Z(\phi)$, and the {\it shape} of $n_{\rm e}(\phi)$
without the knowledge of $d_{\rm A}$ or $d_{\rm L}$ (see
\cite{boehringer10} for a recent review), whereas the absolute value of
$n_{\rm e} (\phi) $ does depend on the distances. Denoting $n_{\rm
e}(\phi) = n_{\rm norm} f_n(\phi)$ to separate the normalization and the
shape ($f_n(\phi)$ is dimensionless and normalized at some scale
radius), one can rewrite equation (\ref{eq-ix1}) as
\begin{eqnarray}
I_{\rm X}(\theta) = \frac{d^3_{\rm A} n^2_{\rm norm}}{d^2_{\rm L}} 
K_{\rm X}(\theta,z),
\label{eq-ix2}
\end{eqnarray}
where observable quantities are  
\begin{eqnarray}
K_{\rm X}(\theta,z)\equiv \frac{1}{2\pi}\int_{\theta}^{\infty}
f_n^2(\phi ) \Lambda_{\rm X}[T_{\rm e}(\phi),Z(\phi),z] 
\frac{\phi d\phi}{\sqrt{\phi^2-\theta^2}}. 
\end{eqnarray}
Similarly, the Compton y-parameter for the same cluster is given by
\begin{eqnarray}
y(\theta) = d_{\rm A} n_{\rm norm} K_{\rm SZ}(\theta), 
\label{eq-iy}
\end{eqnarray}
where 
\begin{eqnarray}
K_{\rm SZ}(\theta)\equiv \frac{2 \sigma_{\rm T} k_{\rm B}}{m_{\rm e}c^2} \int_{\theta}^{\infty}
f_n(\phi ) T_{\rm e}(\phi) \frac{\phi d\phi}{\sqrt{\phi^2-\theta^2}}. 
\label{eq-intsz}
\end{eqnarray}
Eliminating $n_{\rm norm}$ from equations (\ref{eq-ix2}) and (\ref{eq-iy})
gives 
\begin{eqnarray}
d_{\rm A} \eta^2 = \frac{y^2(\theta)}{I_{\rm X}(\theta) (1+z)^4}
\frac{K_{\rm X}(\theta,z)}{K^2_{\rm SZ}(\theta)},  
\label{eq-xsz1}
\end{eqnarray}
where 
\begin{eqnarray}
\eta \equiv \frac{d_{\rm L}}{d_{\rm A} (1+z)^2}  
\label{eq-dual}
\end{eqnarray}
is unity if the distance duality relation holds \cite{uzan04}
\footnote{Equation (\ref{eq-dual}) is different from the definition
adopted in \cite{uzan04} but widely used in more recent studies (e.g.,
\cite{bernardis06,nair11,cardone12,holanda12}).}.  The right hand side of
equation (\ref{eq-xsz1}) consists of observables and gives a direct
measure of $d_{\rm A} \eta^2$. For given $n_{\rm e}$, $T_{\rm e}$, and
$Z$ as a function of the physical radius $d_{\rm A}\phi$, equations
(\ref{eq-ix1}) and (\ref{eq-iy}) indicate that $I_{\rm X} \propto
\eta^{-2} (1+z)^{-4}$ and $y$ is independent of $z$, respectively.

\begin{figure}[tbp]
\centering\includegraphics[width=76mm]{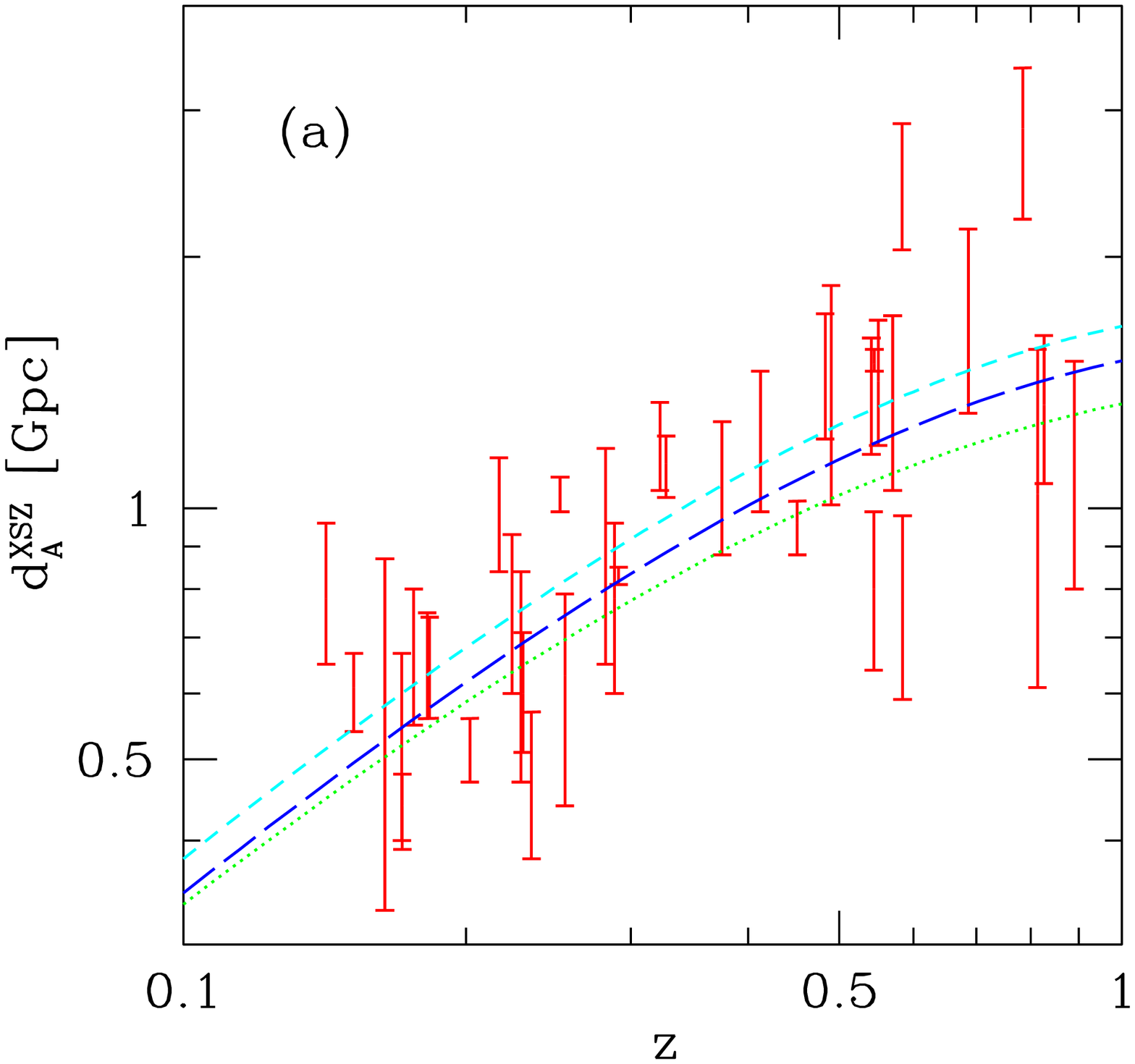}
\centering\includegraphics[width=76mm]{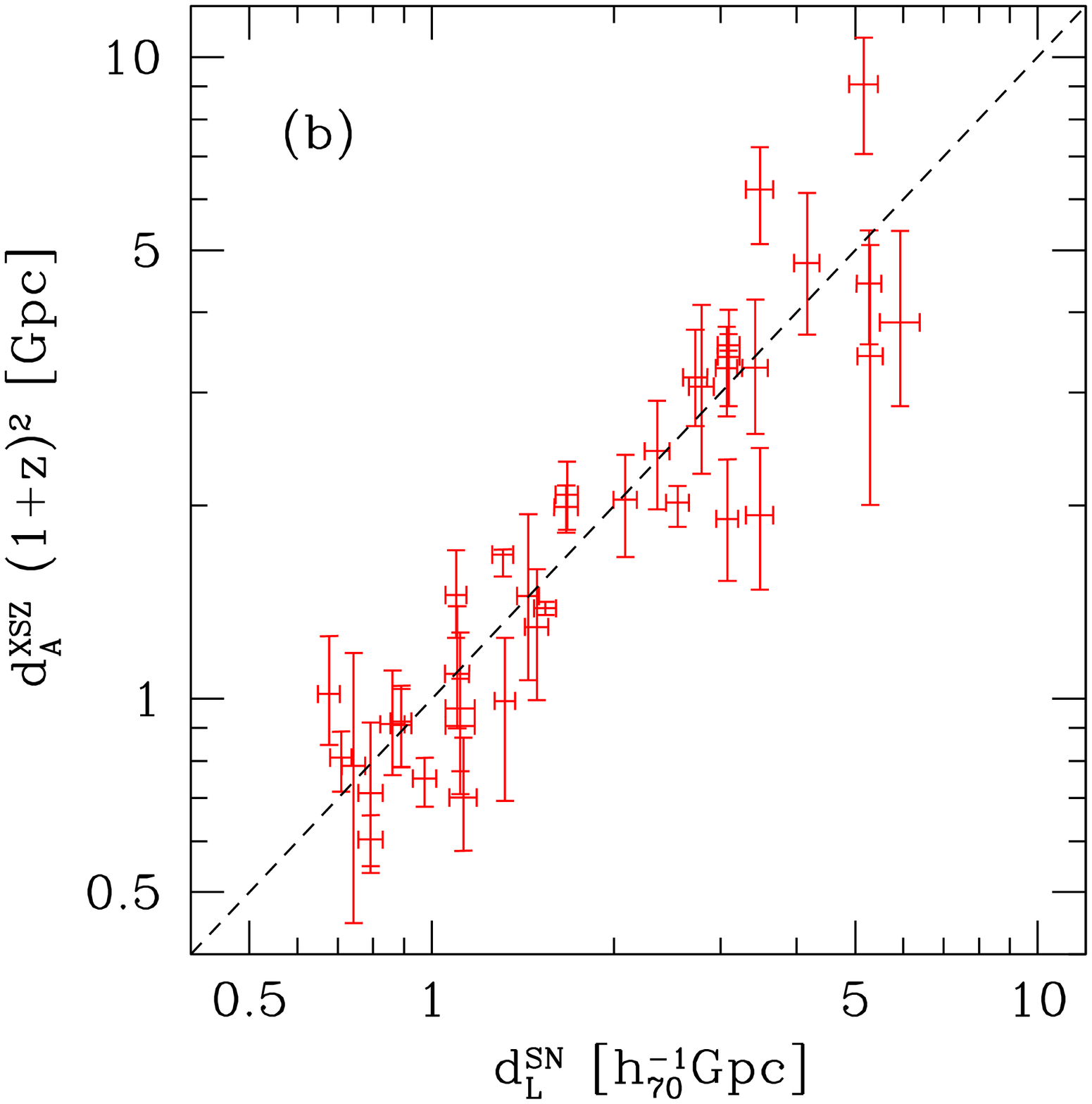} \caption{{\it
Left}: (a) Angular diameter distances of 38 clusters measured by Chandra
X-ray and OVRO/BIMA SZE data versus redshift \cite{bonamente06}. Lines
indicate theoretical predictions for $(\Omega_{\Lambda},h_{70})=
(0.7,1.0)$ (short dashed), $(0.7,1.1)$ (long dashed), and $(0,1.1)$
(dotted) assuming $\Omega_{\rm m}=0.3$, $w=-1$, and $\eta=1$. {\it
Right}: (b) Same quantities multiplied by $(1+z)^2$ versus luminosity
distances measured by the Union2.1 compilation of Type Ia supernova data
\cite{suzuki12}.  The values of $d_{\rm L}^{\rm SN}$ are the mean and
its error of supernova distances whose redshifts match that of a galaxy
cluster within $3\%$; there are on average 10 such supernovae per each
cluster and an error due to the redshift differences is included in the
error of $d_{\rm L}^{\rm SN}$. Dashed line marks the distance duality
relation, $\eta=1$.  } \label{fig-xsz}
\end{figure}

Historically, equation (\ref{eq-xsz1}) has been used widely to measure
the Hubble constant assuming the distance duality relation ($\eta=1$)
and the standard Friedmann-Lemaitre universe, in which case
\begin{eqnarray}
\label{eq-da}
d_{\rm A} &=& \frac{c}{H_0(1+z)}\left\{  \begin{array}{ll}
\frac{\sinh(\sqrt{\Omega_{\rm K}}~\chi)}{\sqrt{\Omega_{\rm K}}} & (\Omega_{\rm K}>0) \\
\chi & (\Omega_{\rm K}=0) \\
\frac{\sin(\sqrt{-\Omega_{\rm K}}~\chi)}{\sqrt{-\Omega_{\rm K}}} & (\Omega_{\rm K}<0) 
\end{array}
\right.
\end{eqnarray}
where $\Omega_{\rm K} = 1 - \Omega_{\rm m} - \Omega_{\Lambda}$ and 
\begin{eqnarray}
\label{eq-chi}
\chi = \int_0^{z} \frac{dz'}{\left[\Omega_{\rm m}(1+z')^3 + \Omega_{\rm
		       K}(1+z')^2 + \Omega_{\Lambda
		       }(1+z')^{3(1+w)}\right]^{1/2}}
\equiv   \int_0^{z} \frac{dz'}{E(z')}.
\label{eq-ez}
\end{eqnarray}
Once the measurements attain sufficient accuracy, one will also be
able to determine $\Omega_{\rm m}$, $\Omega_{\Lambda}$, and $w$
\cite{kobayashi96}. Early measurements assumed that the gas is
isothermal ($T_{\rm e}(\phi)=$\,constant) and tended to yield low values
of $H_0$; e.g., a fit to the ensemble of 38 distance measurements
compiled from the literature gave $H_0 =60 \pm 3$ km\:s$^{-1}$Mpc$^{-1}$
assuming $(\Omega_{\rm m}, \Omega_{\Lambda}, w) =(0.3, 0.7, -1)$
\cite{carlstrom02}. More recent studies, that take account of the
radial variation of $T_{\rm e}$ using spatially resolved X-ray
spectroscopic observations by Chandra, report $H_0=69\pm 8$
km\:s$^{-1}$Mpc$^{-1}$ and $77^{+4+10}_{-3-8}$ km\:s$^{-1}$Mpc$^{-1}$
from 3 clusters at $0.09<z<0.45$ \cite{schmidt04} and 38 clusters at
$0.14<z<0.89$ \cite{bonamente06}, respectively, for $(\Omega_{\rm m},
\Omega_{\Lambda}, w) =(0.3, 0.7, -1)$. As noted by
\cite{schmidt04}, direct temperature measurements were available out to
about one-third of the virial radius for most clusters and the
temperatures at larger radii were estimated assuming that the gas is in
hydrostatic equilibrium with the gravitational potential inferred from
numerical simulations \cite{nfw97}. Figure \ref{fig-xsz}(a) compares
the values of $d_{\rm A}$ measured by \cite{bonamente06} in such
non-isothermal hydrostatic equilibrium model using the OVRO/BIMA SZE
data and a range of theoretical predictions. As discussed later, a large
scatter of the data is partly ascribed to asphericity of clusters and
careful control of systematic errors will be crucial for improving the
accuracy of the distance measurement.

Alternatively, given the knowledge of cosmological parameters including
$H_0$ from other measurements, e.g., CMB primary anisotropies and
Cepheid variables, one can search for any departure from the distance
duality relation using the same sets of data. Denoting the observable on
the right hand side of equation (\ref{eq-xsz1}) by $d_{\rm A}^{\rm
XSZ}$, the quantity $\eta$ is written as
\begin{eqnarray}
 \eta = \sqrt{\frac{d_{\rm A}^{\rm XSZ}}{d_{\rm A}}},
\label{eq-dual2}
\end{eqnarray}
or equivalently from equation (\ref{eq-dual}), 
\begin{eqnarray} \eta = \frac{d_{\rm
A}^{\rm XSZ}}{d_{\rm L}}(1+z)^2.  
\label{eq-dual3}
\end{eqnarray} 
One way of performing a consistency test is to use the predicted values
of $d_{\rm A}$ from equation (\ref{eq-da}) in equation (\ref{eq-dual2});
$\eta$ should be unity over the range of redshifts considered if the
distance duality relation holds and the correct cosmological model is
used for $d_{\rm A}$ \cite{uzan04,bernardis06}.  A more
model-independent test is to use the measured values of $d_{\rm L}$,
e.g., from Type Ia supernovae, in equation (\ref{eq-dual3})
\cite{nair11,cardone12,holanda12}. Figure \ref{fig-xsz}(b) shows $d_{\rm
A}^{\rm XSZ} (1+z)^2$ for 38 galaxy clusters from \cite{bonamente06}
against $d_{\rm L}^{\rm SN}$ from the Union2.1 compilation of the Type
Ia supernova data \cite{suzuki12}.  For the latter, we extract from
publicly available distance
moduli\footnote{http://supernova.lbl.gov/Union/} of 580 supernovae the
mean luminosity distance for those that fall within $|\Delta z|/z <
0.03$ from each of 38 galaxy clusters; the range of $\Delta z$ is chosen
so that its impact on the error of $d_{\rm L}^{\rm SN}$ is comparable to
that from the distance modulus error and on average 10 supernovae are
assigned for each cluster\footnote{For definiteness, the contribution
from $\Delta z$ to $\Delta d_{\rm L}^{\rm SN}$ is computed for
$(\Omega_{\rm m}, \Omega_{\Lambda}, w) =(0.3, 0.7, -1)$ and added in
quadrature to that from the distance modulus error. The resulting
$\Delta d_{\rm L}^{\rm SN}$ is insensitive to the assumed cosmological
parameters.}.  Note that the supernovae only provide relative distance
measurements and $h_{70}=1$ is assumed for determining the absolute
magnitude in the Union2.1 compilation; i.e., $d_{\rm L}^{\rm SN}$
plotted in Figure \ref{fig-xsz}(b) is proportional to $h_{70}^{-1}$.  No
significant deviation from $\eta=1$ has been detected in the current
data out to $z \sim 0.8$. We will hence assume $\eta=1$ in the rest of
this paper unless stated explicitly.

The distance determination by the SZE and X-ray technique is highly
complementary to other astronomical methods and directly applicable to
high redshifts. Controlling various systematic effects is crucial for
improving its accuracy.  First, a departure from spherical symmetry
leads to overestimation of $d_{\rm A}$ (underestimation of $H_0$) if the
cluster is elongated along the line-of-sight and vice versa; as
described in Section \ref{sec-struct}, this property can in turn be used
for studying the gas distribution. While asphericity primarily enhances
the scatter of measurements, there may also be a systematic bias owing
to the fact that such elongated clusters are brighter and easier to
observe; it has been pointed out that strongly elongated clusters are
preferentially aligned along the line-of-sight in a sample of 25 X-ray
selected clusters with existing SZE data
\cite{defilippis05}. Measurements using a homogeneous sample in both
X-rays and SZE will be crucial for eliminating this bias.  Second,
clumpiness in the gas density will reduce $y^2/I_{\rm X} \sim \langle
n_{\rm e} \rangle^2/\langle n_{\rm e}^2 \rangle$ in equation
(\ref{eq-xsz1}) and bias the value of $d_{\rm A}$ low, possibly by
$10 - 20 \%$ \cite{inagaki95}.  On the other hand, inhomogeneities of the
gas temperature give rise to overestimation of $d_{\rm A}$ and may
surpass the bias by the density clumpiness \cite{kawahara08}. Third,
unresolved point sources in the SZE decrement/increment data will
reduce/enhance the estimated value of $d_{\rm A}$.  Finally, calibration
uncertainties of absolute intensities and the temperature in X-ray and
SZE observations are likely to be responsible for additional $10 - 20\%$
errors (e.g., \cite{reese10}).

\subsection{Gas mass fraction of galaxy clusters}
\label{sec-fgas}

Largest clusters of galaxies have grown out of density fluctuations
spread over a comoving scale of $>10$ Mpc and are expected to be fair
samples of the matter content of the Universe. Their baryonic-to-total
mass ratio should therefore provide a measure of $\Omega_{\rm
b}/\Omega_{\rm m}$ \cite{white93}; if a part of baryonic mass in
clusters is observed, a robust lower bound to $\Omega_{\rm
b}/\Omega_{\rm m}$ can still be obtained.  Furthermore, the fact that
the baryon fraction should be constant with redshifts can be used to
measure the distances independently of the absolute value of
$\Omega_{\rm b}/\Omega_{\rm m}$ \cite{sasaki96,pen97} as described
below.

Baryons in clusters are dominated by hot thermal plasma observed
with X-rays and the SZE. The intracluster plasma is almost fully ionized
and close to the primordial composition of hydrogen and helium plus a
small fraction ($<1\%$ in weight) of heavier elements. The gas mass can
therefore be measured by integrating $n_{\rm e}$ over the volume.  From
equations (\ref{eq-ix1}) and (\ref{eq-iy}), $M_{\rm gas} \propto \int
n_{\rm e} dV \propto n_{\rm norm} d^3_{\rm A}$ gives
\begin{eqnarray}
M_{\rm gas, X} \propto d_{\rm L} d_{\rm A}^{3/2},
\label{eq-mgasx}
\end{eqnarray}
and 
\begin{eqnarray}
M_{\rm gas, SZ} \propto  d_{\rm A}^{2},
\label{eq-mgassz}
\end{eqnarray}
for X-rays and the SZE, respectively.

Observations of the intracluster gas further provide a measure of the
total mass enclosed within a physical radius $r=d_{\rm A}\phi$ on
the assumption that the gas is in hydrostatic equilibrium with the
gravitational potential as
\begin{eqnarray}
M(<r) = - \frac{r}{G}\frac{k_{\rm B}T_{\rm e}(r)}{\mu m_{\rm p}}\left[
\frac{d\ln n_{\rm e}(r)}{d\ln r}
+ \frac{d\ln T_{\rm e}(r)}{d\ln r} \right], 
\label{eq-hse}
\end{eqnarray}
where $\mu$ is the mean molecular weight and $m_{\rm p}$ is the proton
mass. Equation (\ref{eq-hse}) does {\it not} depend on the absolute
value of $n_{\rm e}$ (or that of pressure $P_{\rm e} \propto n_{\rm e}
T_{\rm e}$) and scales linearly with the distance as $M \propto d_{\rm
A}$. The total mass can also be estimated using galaxy velocity
dispersions or gravitational lensing with the similar scaling with the
distance to the cluster (strictly speaking, lensing mass depends on the
relative positions of the source and the cluster which introduce an
additional weak cosmological dependence). In practice, the mass can be
measured within a finite radius often expressed in terms of a scaled
radius $R_{\Delta}$, defined as the radius within which the average
matter density is $\Delta$ times the critical density of the Universe;
e.g., $R_{2500}$ and $R_{500}$ correspond to about 20\% and 50\%,
respectively, of the virial radius ($\simeq R_{100}$) at $z=0$ for the
mass profile inferred from numerical simulations \cite{nfw97}.
Likewise, $M_\Delta$ denotes the total mass enclosed within $R_{\Delta}$
and they are related by
\begin{equation}
R_\Delta = 
1.5 \mbox{~Mpc} \left(
\frac{M_\Delta}{10^{15}M_\odot}\right)^{1/3}
\left(\frac{\Delta}{500}\right)^{-1/3} 
h_{70}^{-2/3} E(z)^{-2/3}, 
\label{eq-rdelta}
\end{equation}
where $E(z)$ is defined in equation (\ref{eq-ez}).

Taken together, the gas mass fraction $f_{\rm gas}=M_{\rm gas}/M$ in
clusters depends on the distance as
\begin{eqnarray}
f_{\rm gas, X} \propto  d_{\rm L} d_{\rm A}^{1/2} \propto 
d_{\rm A}^{3/2}(1+z)^2 \eta, 
\label{eq-fgasx}
\end{eqnarray}
and 
\begin{eqnarray}
f_{\rm gas, SZ} \propto  d_{\rm A}, 
\label{eq-fgassz}
\end{eqnarray}
for gas masses measured with X-rays and the SZE, respectively. Equation
(\ref{eq-fgasx}) implies that one can also test the distance duality
relation using $f_{\rm gas, X}$ if it is intrinsically constant over the
range of redshifts observed \cite{goncalves12}.  While $f_{\rm gas, SZ}$
depends on $d_{\rm A}$ more weakly than $f_{\rm gas, X}$, it has an
advantage of being less sensitive to clumpiness of the gas.  As
expected, equating the two quantities, $f_{\rm gas, X}=f_{\rm gas, SZ}$,
recovers essentially the same measure of the distance as equation
(\ref{eq-xsz1}). Obviously, possible evolution of the other baryon
components such as stars and the gas depleted from the clusters is the
major source of systematic errors and needs be properly taken into
account.

For the same set of 38 clusters as the one used for the $H_0$
measurement in \cite{bonamente06}, the inferred gas mass fraction is
consistent with a constant value of $f_{\rm gas, SZ} \simeq 0.12\,
h_{70}^{-1}$ within $R_{\rm 2500}$, albeit with a large scatter, over
$0.14 < z <0.89$ for $(\Omega_{\rm m}, \Omega_{\Lambda}, w) =(0.3, 0.7,
-1)$ \cite{laroque06}.  This is in agreement with independent SZE
measurements by VSA \cite{lancaster05} and AMiBA \cite{umetsu09} as well
as measured values of $f_{\rm gas, X}$
\cite{allen08,vikhlinin09a,ettori09}; it corresponds to $\sim 80 \%$ of
the cosmic mean value $\Omega_{\rm b}/\Omega_{ \rm m}$ from the Planck
2013 results \cite{planck_cosm}.  Observed $f_{\rm gas, X}$ of nearby
clusters tends to increase with the radius from the cluster center
\cite{vikhlinin06,zhang10,simionescu11,eckert13b} and it may partly be
due to clumpiness and substructures. It will hence be meaningful to
improve the sensitivities of the $f_{\rm gas, SZ}$ measurements
particularly at large radii ($> R_{\rm 500}$).  Since the SZE directly
measures the gas mass projected on the sky times the mass-weighted
temperature (eq. [\ref{eq-szflux}]), it can also be combined with the
projected total mass from weak lensing to yield cylindrical $f_{\rm gas,
SZ}$ without an assumption of spherical symmetry \cite{holder00}.

\section{Source Counts}
\label{sec-nz}

The ability to find a galaxy cluster in SZE surveys is primarily limited
by its flux, which is proportional to equation (\ref{eq-iy}) integrated
over the sky,
\begin{eqnarray}
S_{\rm SZ} \propto 
\int y(\theta) d^2 \theta \propto \frac{1}{d^2_{\rm A}} \int n_{\rm e}
 T_{\rm e} dV \propto 
\frac{M_{\rm gas} \langle T_{\rm e} \rangle }{d^2_{\rm A}},
\label{eq-szflux}
\end{eqnarray}
where $\langle \cdots \rangle$ denotes the mass-weighted average.  Since
$d_{\rm A}$ depends on $z$ only weakly at $z>0.5$ and $T_{\rm e}$
correlates with mass, flux-limited SZE surveys become nearly
mass-limited at high redshifts.  The X-ray flux, on the other hand, is
given from equation (\ref{eq-ix1}) as
\begin{eqnarray}
S_{\rm X} = \int I_{\rm X}(\theta) d^2 \theta \propto 
\frac{1}{d^2_{\rm L}} \int n^2_{\rm e}
 \Lambda_{\rm X}(T_{\rm e},Z,z) dV.  
\label{eq-xflux}
\end{eqnarray}
While a rapid increase of $d^2_{\rm L}$ with $z$ is partly canceled by
the evolution of $n_{\rm e}$, finding low-mass clusters becomes more
challenging at higher $z$ in flux-limited X-ray surveys. Typical radius
of a galaxy cluster $\sim$ Mpc (eq. [\ref{eq-rdelta}]) corresponds to
$\sim 2'$ at $z=1$ and angular resolution better than this scale is also
necessary to identify distant clusters. A rapid decrease in the number
of galaxy clusters detected by Planck with redshift shown in Figure
\ref{fig-nsz} is likely due to its moderate spatial resolution of $\ngt
5'$ \cite{planck_hfi}, whereas SPT and ACT are designed for finding
clusters up to high $z$ with beam FWHMs at 150\:GHz of $1.2'$
\cite{spt09} and $1.4'$ \cite{act10}, respectively. On the other hand,
Planck covers a wider frequency range up to $>300$ GHz and is more
suitable for observing nearby clusters including their SZE increment
signals (Fig. \ref{fig-szimage}).

\begin{figure}[tbp]
\centering\includegraphics[width=76mm]{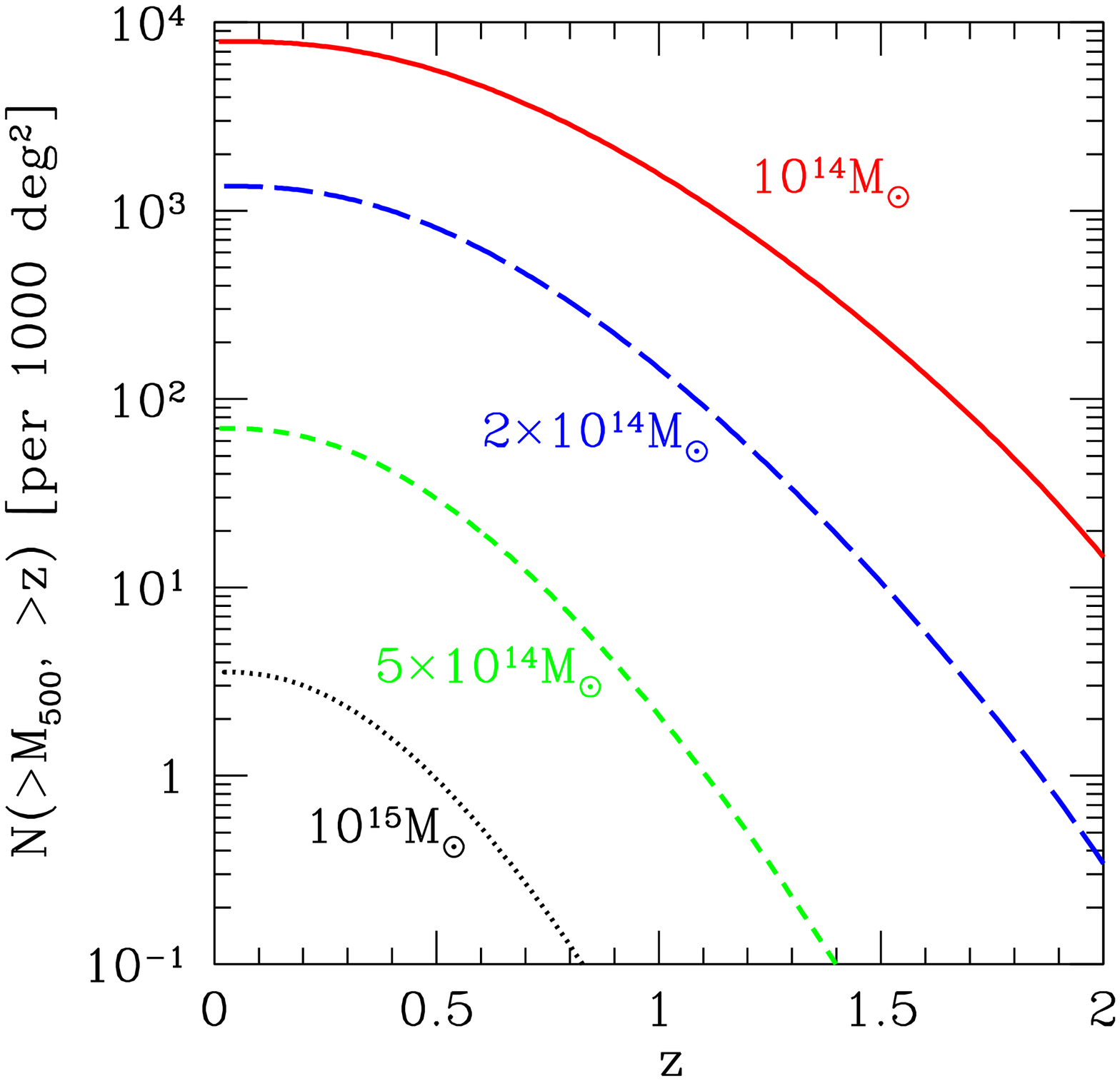}
\centering\includegraphics[width=76mm]{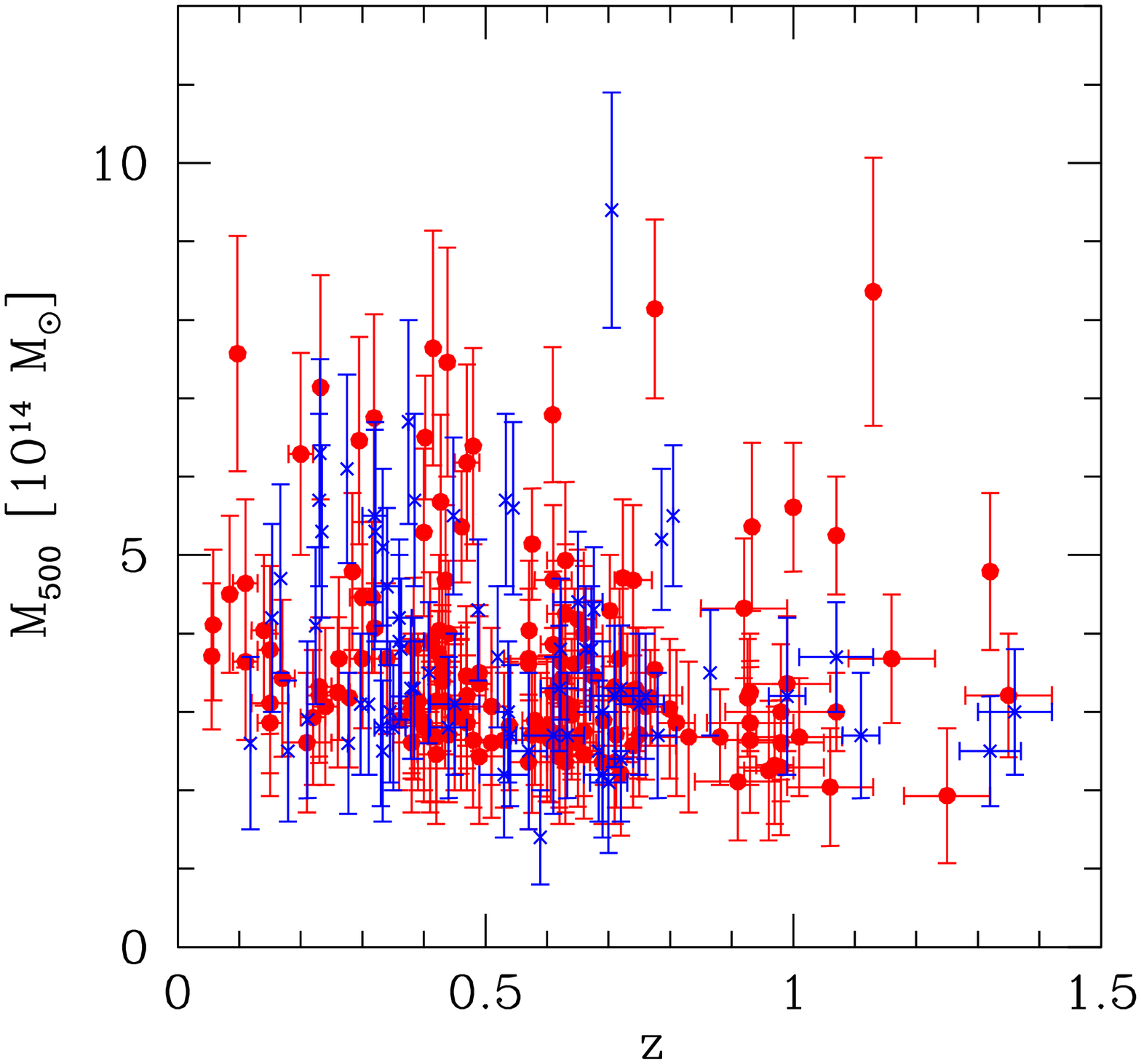} \caption{{\it
Left}: Predicted numbers of galaxy clusters per 1000 deg$^2$ above given
redshift $z$ and mass $M_{500}$. The numbers are plotted for $M_{500} =
10^{14} M_\odot$ (solid), $2\times 10^{14} M_\odot$ (long dashed),
$5\times 10^{14} M_\odot$ (short dashed), and $10^{15} M_\odot$
(dotted), using the mass function of \cite{tinker08} and assuming the
conventional $\Lambda$CDM universe. {\it Right}: Estimated masses versus
redshifts of galaxy clusters detected in the surveys by SPT in 720
deg$^2$ (red circles, \cite{spt13a}) and ACT in 504 deg$^2$ (blue
crosses, \cite{act13a}). The plotted are 158 and 68 clusters confirmed
by optical/infrared imaging and include 117 and 19 new discoveries,
respectively.}  \label{fig-nz}
\end{figure}

The expected number of sources per unit solid angle above the flux $S$
between redshifts $z_{\rm min}$ and $z_{\rm max}$ can be written as
(e.g., \cite{kss98}) 
\begin{eqnarray}
N(>S, z_{\rm min}, z_{\rm max}) = \int_{z_{\rm min}}^{z_{\rm max}}dz
\frac{dV}{d\Omega dz} \left.\int_S^\infty dS'\frac{dn(M,z)}{dM}
\frac{dM}{dS'}\right|_{M=M(S',z)}, 
\end{eqnarray}
where $dn(M,z)$ is the comoving number density of galaxy clusters of
mass $M \sim M+dM$ corresponding to flux $S'\sim S'+dS'$ at $z$ and
\begin{eqnarray}
\frac{dV}{d\Omega dz} 
= \frac{c}{H_0}\frac{(1+z)^2 d_{\rm A}^2(z)}{E(z)}
\label{eq-dvdz}
\end{eqnarray}
is the comoving volume element per unit solid angle and unit
redshift. For given initial distribution and evolution thereafter of
primordial density fluctuations, the mass function $dn(M,z)/dM$ can be
computed using either analytic prescriptions (e.g.,
\cite{ps74,sheth99}) or state-of-the-art numerical simulations (e.g.,
\cite{jenkins01,tinker08}), on the assumption that every virialized dark
matter halo above some threshold mass becomes a galaxy cluster.  The
relation between the observed flux and the mass is often estimated by
means of empirical scaling relations calibrated by local observations
(e.g., \cite{vikhlinin09a,arnaud10}). Comparisons with observed numbers
of clusters then give a measure of cosmological parameters through both
the growth of density fluctuations and the geometry of the Universe.

Figure \ref{fig-nz} illustrates the number counts of galaxy clusters
predicted in the conventional $\Lambda$CDM model as well as estimated
masses versus redshifts of clusters detected in the SZE surveys.
Current surveys by SPT and ACT have been finding clusters down to
$M_{500} \simeq 2 \times 10^{14} M_\odot$ up to $z\sim 1.5$ over the
fields of nearly 1000 square degrees \cite{spt13a,act13a}. It should be
noted that completeness of the samples degrades toward low mass and the
estimated masses may be biased particularly at higher $z$ since they are
based on empirical relations extrapolated from low $z$. Within such
uncertainties, the detected numbers are consistent with the predictions
and it is likely that one will start to find clusters at $z>2$ by
reaching deeper fluxes corresponding to $M_{500} < 10^{14} M_\odot$.

The predicted numbers of clusters are the most sensitive to underlying
values of $\Omega_{\rm m}$ and $\sigma_8$. Recent results using a sample
of 189 clusters from the Planck SZE catalog indicate $\sigma_8
(\Omega_{\rm m}/0.27)^{0.3} = 0.764 \pm 0.025$
\cite{planck_counts}. This is in agreement with other measurements in
the local Universe using SZE cluster counts by SPT
\cite{benson13,spt13a} and ACT \cite{act13a}, X-ray cluster counts
\cite{vikhlinin09b} and cosmic shear \cite{kilbinger13}, whereas it
tends to be smaller than that inferred from CMB primary anisotropies
measured by Planck \cite{planck_cosm}. The origin of this tension is
still not entirely clear but may be ascribed to incomplete instrumental
calibration, underestimating true masses of clusters, missing a fraction
of massive clusters, suppression of density fluctuations at small scales
by, e.g., massive neutrinos, or any combination thereof.

Since clusters of galaxies comprise the largest virialized structures in
the Universe, the evolution of their numbers up to high $z$ provides a
sensitive probe of the growth of cosmic structures, which is highly
complementary to purely geometrical methods such as Type Ia supernovae
and Baryon Acoustic Oscillations. It can be used to explore the nature
of dark energy within a framework of standard cosmology (e.g.,
\cite{vikhlinin09b,mantz10}) as well as to search for any departure from
the standard framework itself. For instance, the linear growth rate of
density fluctuations $D$ can be generalized as \cite{wang98,linder05}
\begin{equation}
\frac{d\ln D}{d\ln a} = \Omega_{\rm m}(a)^{\gamma_{\rm g}}
,
\end{equation}
where $a=1/(1+z)$ is the cosmic scale factor, $\Omega_{\rm
m}(a)=\Omega_{\rm m} a^{-3}E^{-2}(a)$, and the index $\gamma_{\rm g}$
takes nearly a constant value $\simeq 0.55$ if general relativity
holds. This will allow one to constrain the growth of structures and the
geometry of the Universe separately from the data.  Current X-ray
cluster data are fully consistent with general relativity
\cite{rapetti13} and the analysis can be refined further by including
higher $z$ clusters such as those observed by the SZE.  

It should be noted that the applicability of cluster counts as a
cosmological probe relies critically on the accuracy of mass
determination. This is a challenging issue particularly at $z> 1$, where
spatially resolved X-ray spectroscopy or weak lensing becomes
increasingly difficult; even if empirical scaling relations are to be
used, they must be calibrated by some independent means. To this end,
SZE imaging observations will further offer a useful measure of the mass
as described in Section \ref{sec-struct}.

\section{Structure of Intracluster Plasma}
\label{sec-struct}

The accuracy of cosmological studies using the SZE is largely limited by
our understanding of astrophysics of galaxy clusters. Historically,
internal structure of the intracluster plasma has been studied
extensively by X-ray observations.  As mentioned in Section
\ref{sec-xsz}, radial profiles of $n_{\rm e}$ and $T_{\rm e}$ have been
measured by X-ray surface brightness and spectra for a large number of
clusters. Modeling the SZE brightness using equation (\ref{eq-iy}) or the
total mass using equation (\ref{eq-hse}) also relied on these
measurements for decades. Detailed X-ray spectroscopic observations,
however, become progressively difficult for distant clusters or toward
the outskirts of even nearby clusters (see \cite{reiprich13} for a
review), owing to low photon counts and background contamination.
Recent developments of high sensitivity and high resolution SZE
observations have opened up new possibilities as described below.

\begin{figure}[tbp]
\centering\includegraphics[width=152mm]{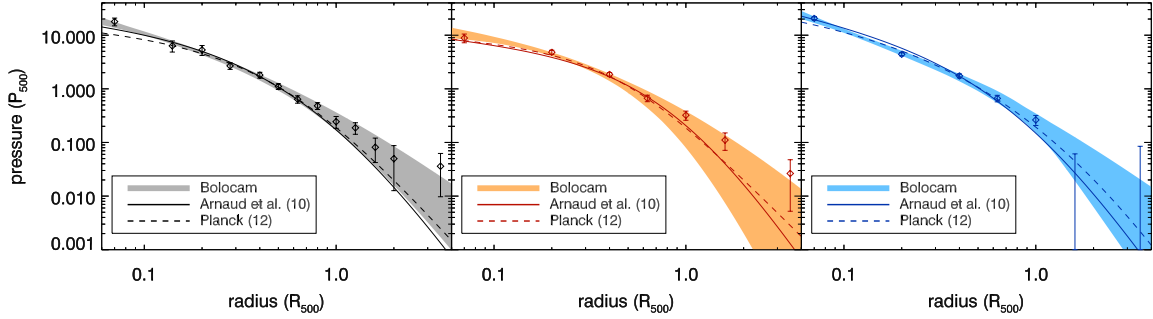}
 \caption{Data points show radial pressure profiles from the Bolocam SZE
 data for 45 clusters at $0.15<z<0.89$ assuming $(\Omega_{\rm m},
 \Omega_{\Lambda}, w, h_{70}) =(0.3, 0.7, -1, 1)$ (reproduced from
 \cite{sayers13a} with permission, \copyright\,AAS); full sample (left panel), a
 subsample of disturbed clusters (middle panel), and a subsample of
 cool-core (apparently relaxed) clusters (right panel). Also shown are
 the fits by a parametric model of equation (\ref{eq-gnfw}) for this
 sample (shaded regions indicating the 68.3\% confidence region), a
 sample of 33 X-ray clusters at $z<0.2$ by Arnaud et al. \cite{arnaud10}
 (solid line), and a sample of 62 Planck-selected clusters at $z\sim
 0.15$ \cite{planck_intv} (dashed line).}  \label{fig-pr}
\end{figure}

First of all, a spatially resolved thermal SZE image alone yields the
radial profile of electron pressure under the assumption of spherical
symmetry. Figure \ref{fig-pr} shows deprojected pressure profiles from
the SZE data taken by Bolocam \cite{sayers13a}.  Radially averaged
pressure at $0.2 R_{500} < r< R_{500}$ is rather insensitive to
dynamical status of clusters and is well represented by the following
functional form \cite{nagai07,arnaud10},
\begin{equation}
 \frac{P_{\rm e}}{P_{500}} = \frac{P_0}{(C_{500}X)^{\alpha_3} [1+
(C_{500}X)^{\alpha_1}]^{(\alpha_2-\alpha_3)/\alpha_1}},
\label{eq-gnfw}
\end{equation}
where $X=r/R_{500}$, $(P_0,C_{500},\alpha_1,\alpha_2,\alpha_3)$ are
fitting parameters, and $P_{500}$ is the scaled pressure defined by
\begin{equation}
P_{500}= 1.65 {\rm ~eV~cm}^{-3} \left(
\frac{M_{500}}{3\times
10^{14} h_{70}^{-1} M_\odot}
\right)^{\beta_{\rm P}} h_{70}^2 
E(z)^{8/3}. 
\label{eq-p500}
\end{equation}
Apart from the fact that slightly different values of $\beta_{\rm P}$
are used in the literature (e.g., 0.67 in \cite{sayers13a} and 0.79 in
\cite{arnaud10,planck_intv}), the above pressure profile accounts for
the X-ray data of nearby clusters \cite{arnaud10} as well as the SZE
data by SPT \cite{plagge10}, CARMA \cite{bonamente12}, and Planck
\cite{planck_intv}. Discrepant results, on the other hand, are reported
on three individual clusters between AMI and Planck \cite{planck_ami},
suggesting a presence of yet unaccounted for systematic effects.
Dispersion of the reconstructed pressure profile provides a key
consistency check of the applicability of the mass estimation assuming
hydrostatic equilibrium (eq. [\ref{eq-hse}]) or any empirical scaling
relations based on it.

\begin{figure}[t]
\centering\includegraphics[width=76mm]{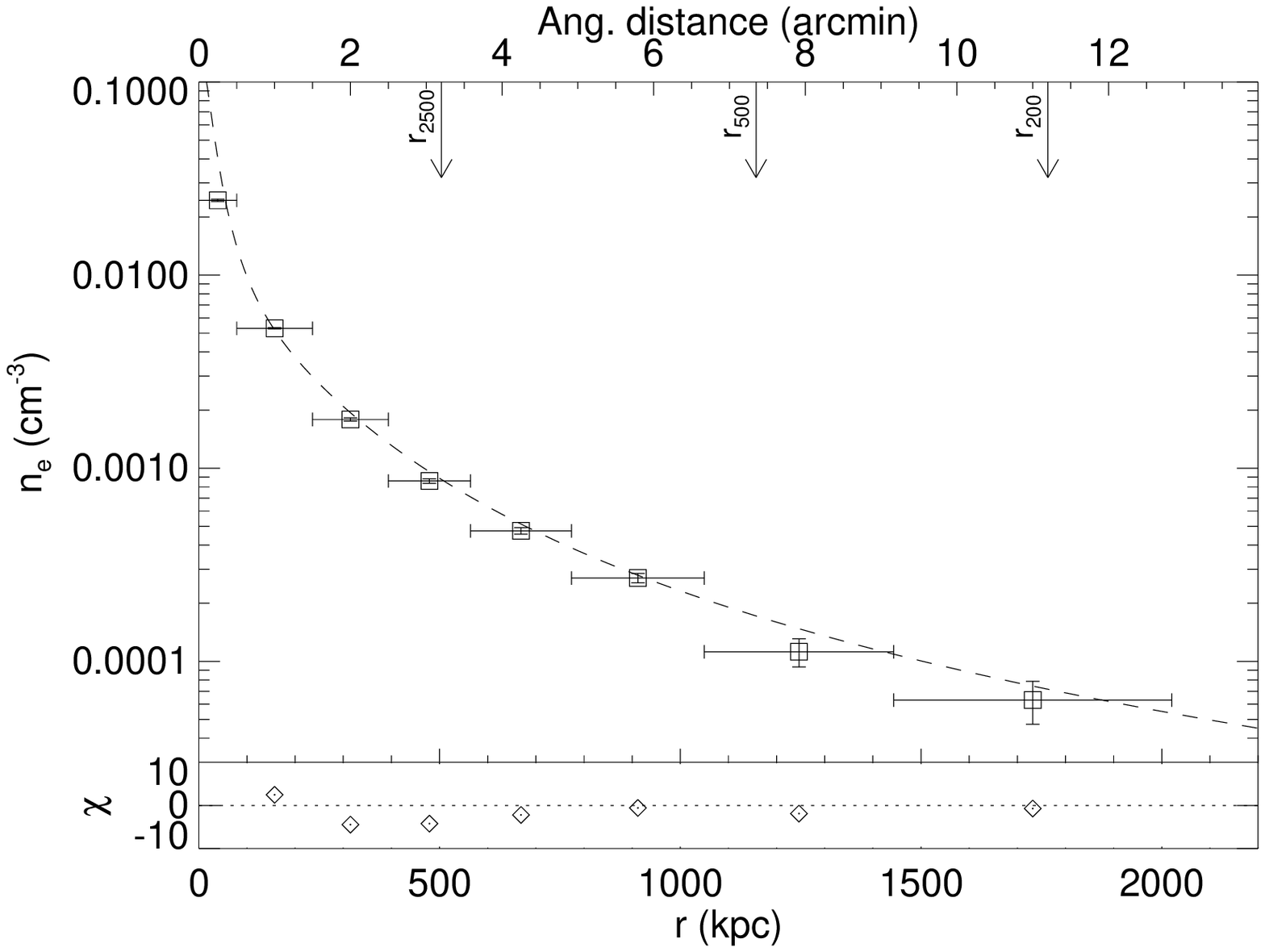}
\centering\includegraphics[width=76mm]{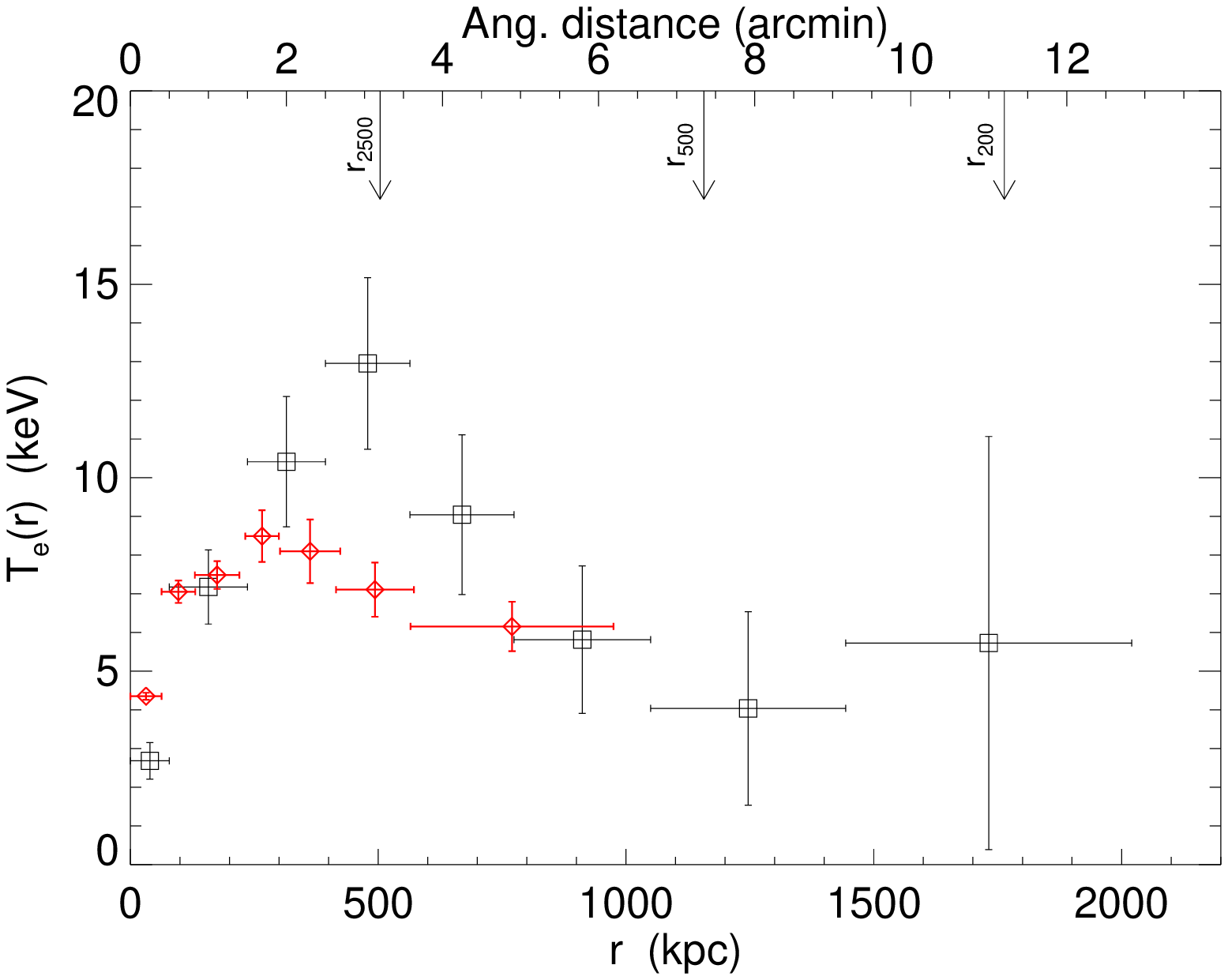}
\caption{Deprojected electron density (left panel) and temperature
(right panel) of Abell 2204 at $z=0.152$ from a joint analysis of the
APEX-SZ SZE image and the XMM-Newton X-ray brightness (reproduced from
\cite{basu10} with permission, \copyright\,ESO). For reference, the dashed line
and red diamonds show the results of X-ray analysis using the spectral
data \cite{zhang08}.}  \label{fig-tr}
\end{figure}

Second, one can combine the SZE image with the X-ray surface brightness
map to recover radial profiles of $n_{\rm e}$ and $T_{\rm e}$ separately
{\it without} X-ray spectroscopic data. This is done essentially by
inverting equations (\ref{eq-ix1}) and (\ref{eq-iy}) using the Abel
transform \cite{silk78,yoshikawa99},
\begin{eqnarray}
n_{\rm e}(\phi)^2 \Lambda_{\rm X}[T_{\rm e}(\phi),Z(\phi),z]
&=& \frac{4 d^2_{\rm L}}{d^3_{\rm A}}
\int_{\phi}^{\infty} \left[-
\frac{dI_{\rm X}(\theta)}{d\theta}\right]
\frac{d\theta}{\sqrt{\theta^2 -  \phi^2}},\\
n_{\rm e}(\phi) T_{\rm e}(\phi)
&=& \frac{m_{\rm e} c^2}{\pi k_{\rm B} \sigma_{\rm T} d_{\rm A}}
\int_{\phi}^{\infty} \left[-\frac{d y(\theta)}{d\theta}\right]
\frac{d\theta}{\sqrt{\theta^2 -  \phi^2}}, 
\label{eq-invsz}
\end{eqnarray}
and separating $n_{\rm e}(\phi)$ and $T_{\rm e}(\phi)$; $\Lambda_{\rm
X}$ depends only weakly on $Z$ for $T_{\rm e} \ngt 2 \times 10^7$ K.
Note that an assumption on underlying cosmology is necessary only to
determine an absolute value of $n_{\rm e}$ or $T_{\rm e}$ and not to
reconstruct the shape of their profiles.  Practical applications of the
above inversion have become possible during the last decade
\cite{kitayama04,yuan08, nord09, basu10, eckert13}. A great advantage of
this method is that it is applicable to X-ray faint regions as long as
imaging data are available; its feasibility has been tested against
existing X-ray spectroscopic measurements as illustrated in Figure
\ref{fig-tr}.  Further invoking an assumption of hydrostatic equilibrium
(eq. [\ref{eq-hse}]), it offers a unique measure of the gravitational
mass at high redshifts and large radii.

\begin{figure}[t]
\centering\includegraphics[width=76mm]{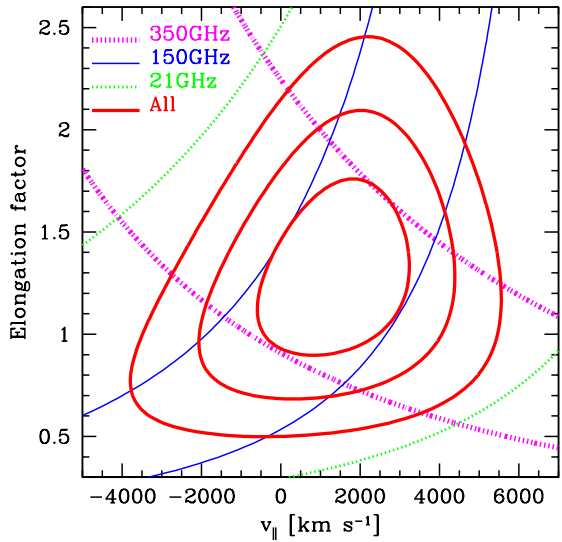}
 \caption{Limits on the line-of-sight elongation factor and the peculiar
 velocity of RX J1347.5--1145 from multi-frequency SZE data (reproduced
 from \cite{kitayama04} with permission, \copyright\,ASJ); the cluster
 is modeled by a spheroid elongated along the line-of-sight. The thick
 solid contours indicate the 68.3, 95.4, and 99.7\% confidence regions
 from a joint fit to the data at three frequencies. The other contours
 show the 68.3\% confidence region from each of the 350 GHz (thick
 dotted), 150 GHz (thin solid), and 21 GHz (thin dotted) data
 separately. The 150 GHz image of this cluster is shown in
 Fig. \ref{fig-rxj1347} and the disturbed substructure is excluded in
 the analysis shown here.}  \label{fig-rxjeta}
\end{figure}

Finally, if an independent measure of $T_{\rm e}$ is also available
through X-ray spectroscopy, one can relax the assumption of spherical
symmetry and explore intrinsic shapes of galaxy clusters for a given
cosmological model. This is in fact an alternative to the distance
determination described in Section \ref{sec-xsz}; a cluster elongated by
some fraction over the line-of-sight will enhance the value of $d_{\rm
A}$ in equation (\ref{eq-xsz1}) by the same fraction. Observed X-ray
images of galaxy clusters have projected axis ratios with a mean $\simeq
0.8$ and a dispersion $\simeq 0.1$ \cite{defilippis05,kawahara10}, and
the SZE data will further add line-of-sight information.  For example,
X-ray and multi-frequency SZE data of Abell 1689 can be explained well
by a mildly triaxial cluster with a minor to major axis ratio of $0.7
\pm 0.15$, preferentially elongated along the line of sight
\cite{sereno12}.  One can further explore the 3D orientation of the dark
matter halo by combining weak lensing data and assuming, for instance,
that the gas is in hydrostatic equilibrium \cite{morandi11} or it shares
the same axis directions with the dark matter \cite{sereno13}. Note that
the SZE brightness of an individual cluster at a single frequency can be
biased by the kinematic SZE; Figure \ref{fig-rxjeta} illustrates that a
combination of multi-frequency data, particularly of both decrement
($\nu <218$ GHz) and increment ($\nu > 218$ GHz) of the SZE, is useful
for breaking the degeneracy between the line-of-sight elongation and the
peculiar velocity.

\section{Dynamics of Galaxy Clusters}

Clusters of galaxies often display signatures of violent mergers, which
comprise the most energetic phenomena in the Universe with the total
kinematic energy $\sim 10^{64}$ ergs and mark directly the sites of
cosmic structure formation. Associations with synchrotron emission from
nonthermal electrons indicate that a certain degree of particle
acceleration is also induced during cluster mergers, although the
precise mechanism is still unknown. While X-ray and low-frequency
($\nlt$GHz) radio observations have been widely used to find such merger
shocks at low redshifts (see \cite{markevitch07,feretti12} for reviews),
the SZE provides a promising and complementary diagnostics up to high
redshifts.  Since the thermal SZE and the kinematic SZE are proportional
to thermal pressure and the bulk velocity, respectively, they serve as
direct probes of shock fronts (i.e., pressure gaps) and gas dynamics. In
addition, SZE images with a spatial resolution of $\sim 10''$
\cite{komatsu01,mason10} or better will continue to play a unique role
in resolving the shock-heated gas with $k_{\rm B}T_{\rm e} \gg 10$ keV,
given that spatial resolutions of current and near future hard X-ray
($E>10$ keV) instruments are limited to $>45''$. By measuring a gap
across the shock of either density, temperature, or pressure, one can
infer the Mach number $\mathcal{M}$ (i.e., gas velocity normalized by
its sound speed in the rest frame of the shock front) from
Rankine-Hugoniot relations:
\begin{eqnarray}
\label{eq-rankine}
  \frac{n_{\rm 2}}{n_{\rm 1}} = \frac{v_1}{v_2} = \frac{4
  \mathcal{M}_1^2}{\mathcal{M}_1^2+3}, ~~~~
\frac{T_{\rm 2}}{T_{\rm 1}} = \frac{(5 \mathcal{M}_1^2
 -1)(\mathcal{M}_1^2 +3)}{16\mathcal{M}_1^2},
\end{eqnarray}
where the subscripts 1 and 2 denote preshock and postshock quantities
respectively, and an adiabatic index of $\gamma=5/3$ has been used. The
product of these equations readily
yield the pressure ratio.

\begin{figure}[t]
\begin{minipage}{0.67\textwidth}
\resizebox{10.4cm}{!}{\includegraphics{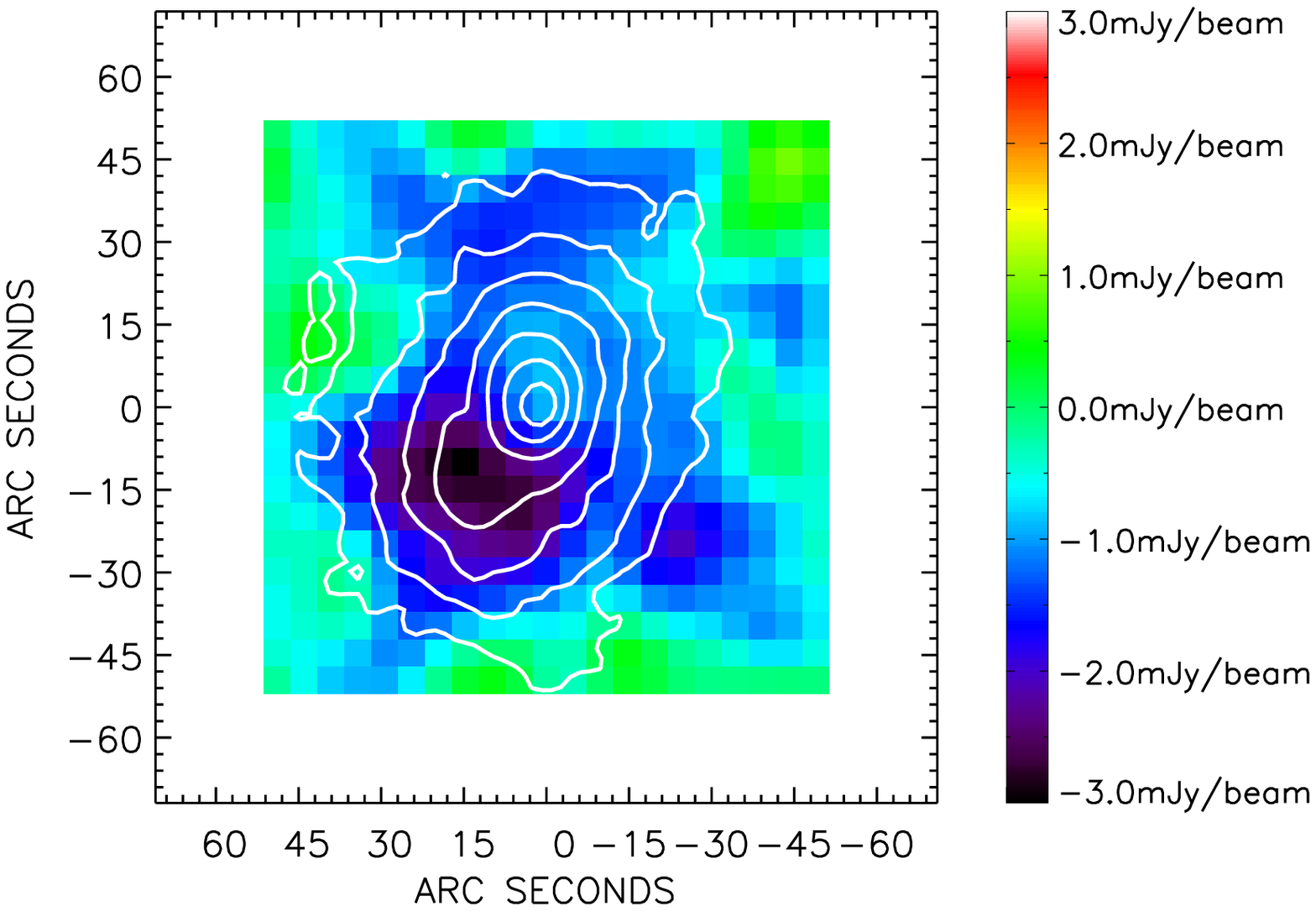}}
\end{minipage}
\begin{minipage}{0.33\textwidth}
\resizebox{4.8cm}{!}{\includegraphics{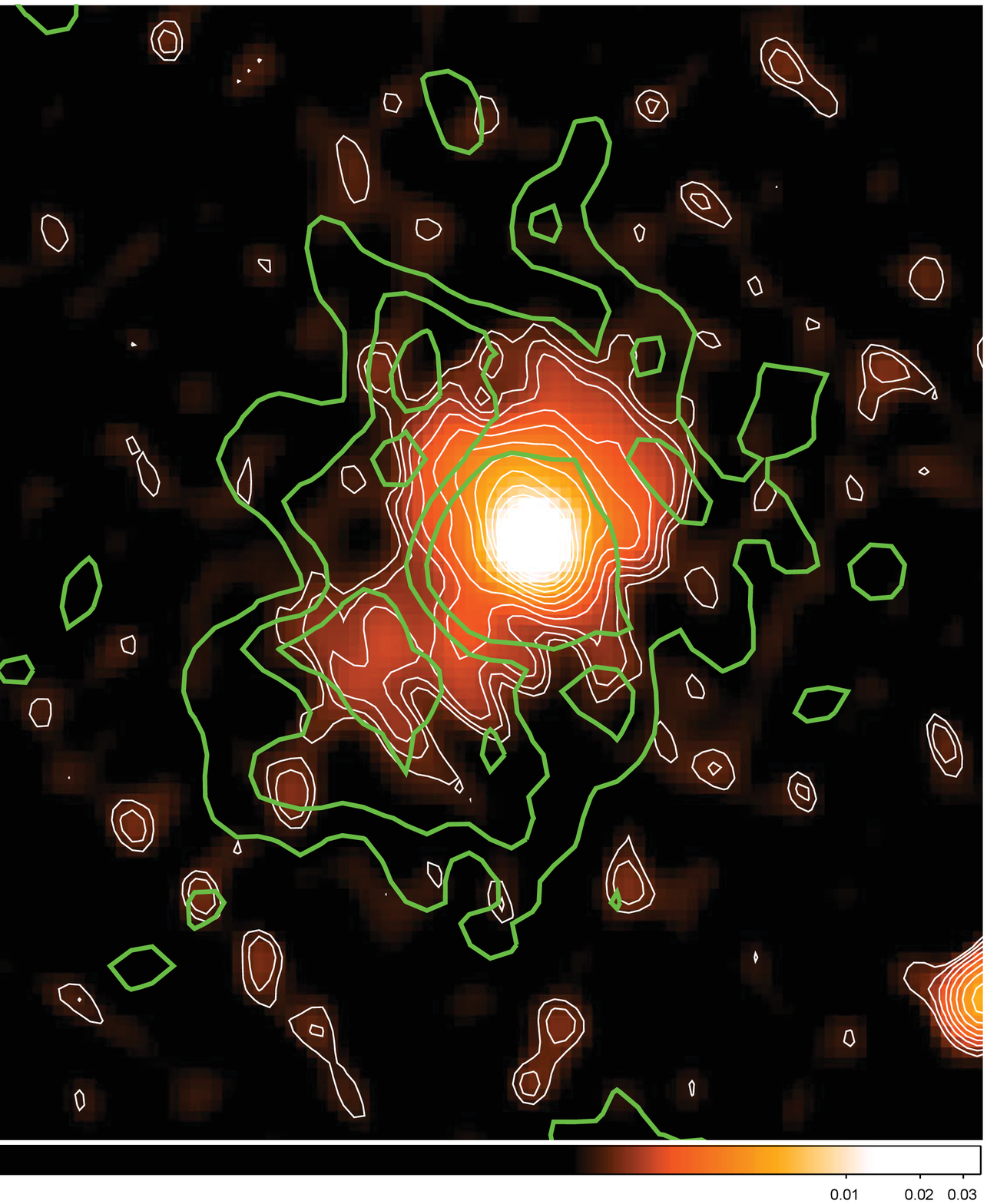}}
\end{minipage}
\caption{Observed images of RX J1347.5-1145 at $z=0.451$; $1''=5.8
 \,h_{70}^{-1}$ kpc in the cluster rest frame. {\it Left:} NOBA 150\:GHz
 SZE map with $13''$ beam FWHM smoothed by a $15''$ Gaussian filter for
 display \cite{komatsu01}, overlaid with Chandra 0.5--7keV X-ray
 brightness contours (reproduced from \cite{kitayama04} with permission,
 \copyright\,ASJ). {\it Right:} GMRT 614\:MHz synchrotron intensity map
 and contours (white), overlaid with MUSTANG 90\:GHz SZE contours with
 $9''$ beam FWHM \cite{mason10} (reproduced from \cite{ferrari11} with
 permission, \copyright\,ESO).  There is a radio point source at the
 cluster center in both images, which {\it reduces} the thermal SZE
 decrements and {\it enhances} the synchrotron intensities.}
 \label{fig-rxj1347}
\end{figure}

A prototype of intensive SZE studies on a merging cluster is given by
those on RX J1347.5-1145 at $z=0.451$, the brightest cluster known to
date in the SZE. This cluster was originally thought to be highly
relaxed, based on smooth morphology of the soft X-ray ($E<2$ keV) image
by ROSAT \cite{schindler97}. The SZE observation by NOBA with $13''$
beam \cite{komatsu01}, however, revealed that it has a prominent
substructure at $\sim 20''$ southeast of the cluster center as shown in
Figure \ref{fig-rxj1347}. This finding has been confirmed subsequently
with Chandra 0.5--7 keV data \cite{allen02} as well as more recent high
sensitivity SZE images by MUSTANG with $9''$ beam
\cite{mason10,korngut11} and by CARMA with the smallest synthesized beam
of $11''\times 17''$ \cite{plagge13}. Independent SZE measurements of
this cluster have also been published using SCUBA
\cite{komatsu99,kitayama04,zemcov07}, Diabolo
\cite{pointecouteau99,pointecouteau01}, OVRO/BIMA
\cite{reese02,carlstrom02}, SuZIE \cite{benson04}, Bolocam and Z-Spec
\cite{zemcov12}.  The inferred temperature of the substructure is
$k_{\rm B}T_{\rm e} \simeq 25$ keV, which is about a factor of $2$
higher than the mean temperature of this cluster $k_{\rm B}T_{\rm e}
\simeq 13$ keV \cite{kitayama04,ota08}; this is in accord with the fact
that the substructure is more obvious in the SZE than soft X-rays. It
follows that the cluster is probably undergoing a major merger; applying
equation (\ref{eq-rankine}) to the above mentioned temperatures gives
the Mach number of $\mathcal{M}_1 \simeq 1.9$ and the corresponding
pre-shock velocity of $v_1 \simeq 3500$ km\:s$^{-1}$.  Figure
\ref{fig-rxj1347} further illustrates that diffuse synchrotron emission
from non-thermal electrons is spatially associated with the hot
substructure \cite{gitti07,ferrari11}.

Merger shocks have also been detected using the SZE for other clusters
including MACS0744.8+3927 at $z=0.69$ with an inferred value of the Mach
number $\mathcal{M}_1 \simeq 1.2$ \cite{korngut11} and Coma at $z=0.023$
with $\mathcal{M}_1 \simeq 2.0$ \cite{planck_coma}. The fraction of
merging clusters is likely to increase with redshifts as the growth of
density fluctuations becomes faster prior to the onset of cosmic
acceleration. It is likely that the SZE surveys will continue to find a
number of new merging clusters as demonstrated by a discovery of ACT-CL
J0102--4915 at $z=0.87$ \cite{menanteau12}.

The kinematic SZE also gives a direct probe of the gas velocity. The
measurement is in general challenging for individual clusters (e.g.,
\cite{zemcov12}), but becomes feasible if a major merger is taking place
along the line-of-sight and high quality SZE data are available at
multi-frequencies. In fact, the line-of-sight velocity of $v_{\parallel}
= -3450 \pm 900$ km\:s$^{-1}$ has been reported for a subcluster of MACS
J0717.5+3745 at $z = 0.55$ using 140\:GHz and 268\:GHz Bolocam data
\cite{sayers13b}. Such measurements are highly complementary to future
high-dispersion X-ray spectroscopic observations using
micro-calorimeters on board
ASTRO-H\footnote{http://astro-h.isas.jaxa.jp/en/} and
ATHENA\footnote{http://www.the-athena-x-ray-observatory.eu/}.

\begin{figure}[t]
\centering\includegraphics[width=77mm]{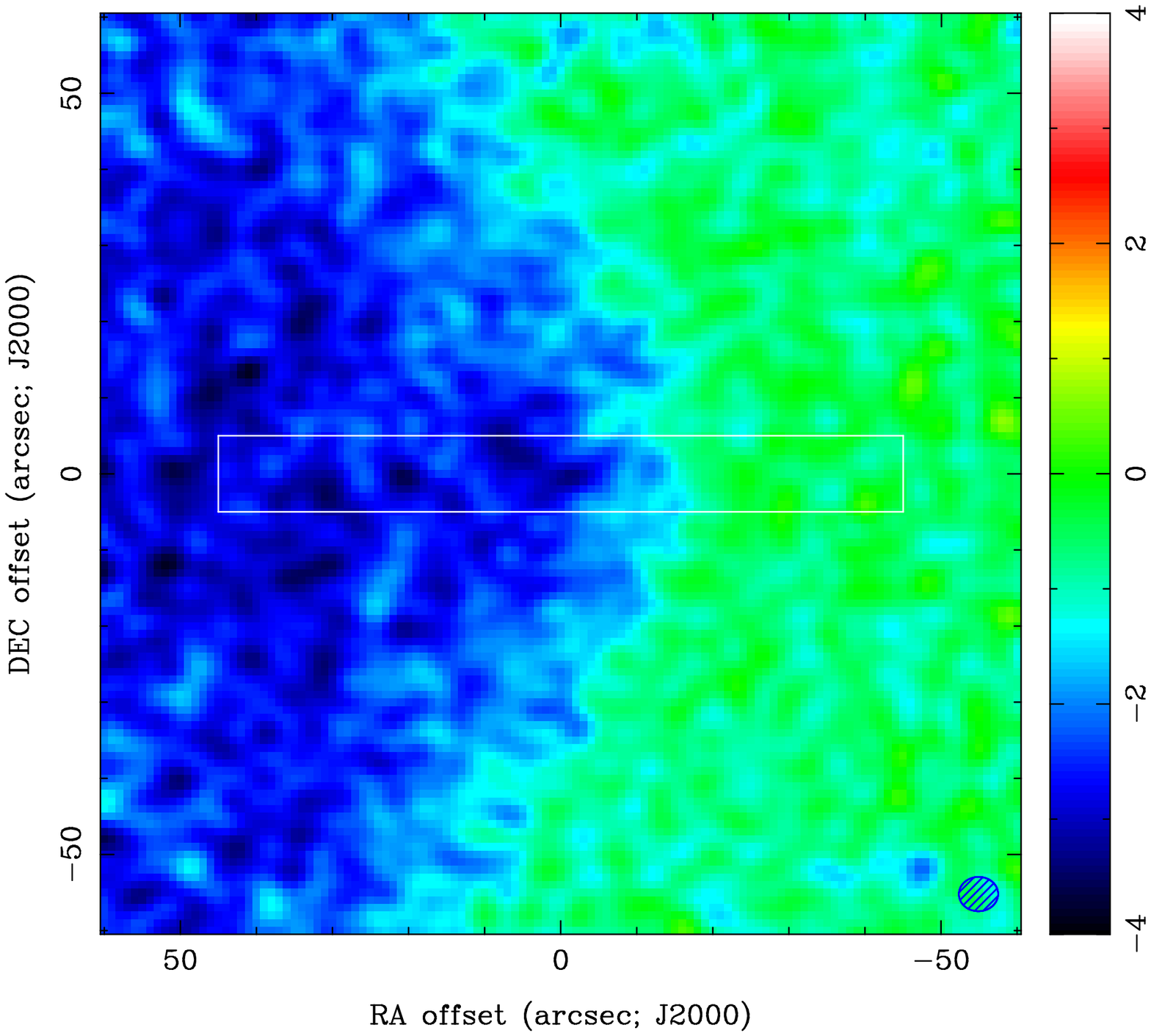}
\centering\includegraphics[width=73mm]{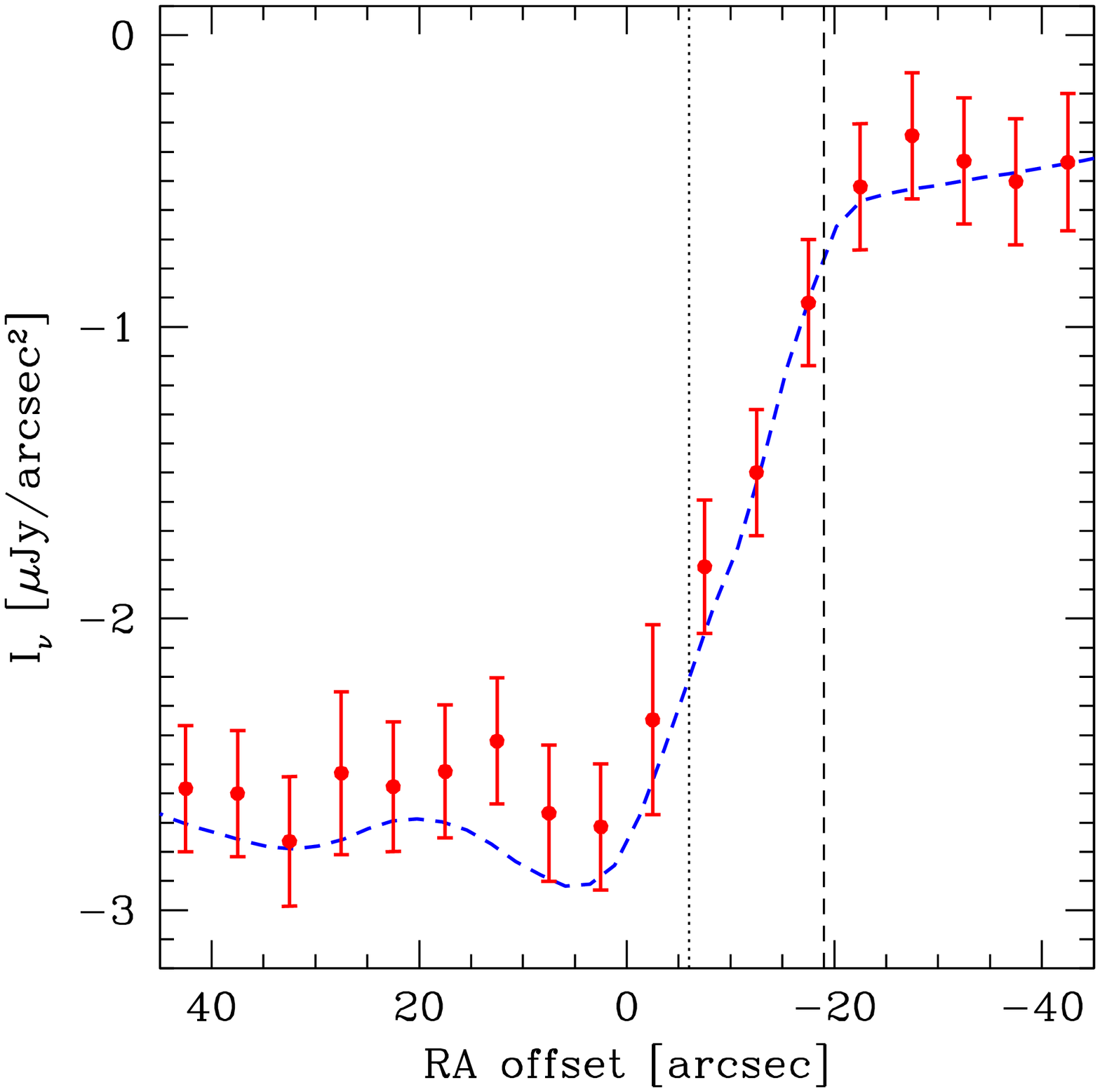}
\caption{Simulations of the shock front in 1E 0657-558 (Bullet cluster)
at $z=0.296$ (reproduced from \cite{yamada12} with permission, \copyright\,ASJ);
$1''=4.4 \,h_{70}^{-1}$kpc in the cluster rest frame. {\it Left}: Expected SZE image
by ALMA at 90GHz including various noise components. The color units are
in $\mu$Jy/arcsec$^2$. {\it Right}: Error bars are the expected SZE
intensities in the region marked by a white box in the left panel,
whereas thick dashed line is the input hydrodynamical model before
adding noise.  Vertical thin lines indicate the positions of the shock
front (thin dashed) and the contact discontinuity (thin dotted),
respectively.
}
\label{fig-shocksect}
\end{figure}

In near future, ALMA will be capable of imaging the SZE in bright
compact clusters with the spatial resolution of $5''$ or better
\cite{yamada12}. Figure \ref{fig-shocksect} demonstrates that ALMA is
indeed a powerful tool for resolving the shock front, characterized by
temperature and pressure jumps.  Shocks in galaxy clusters are in
general hard to find in X-rays because they often appear at outskirts
and are also hidden by a sharp radial gradient of $n_{\rm e}^2$; density
peaks behind the contact discontinuity are much easier to be seen in
X-rays as cold fronts (e.g., \cite{markevitch07}). The SZE and X-rays
are thus complementary in probing the detailed shock structure and the
former is particularly useful for detecting hot rarefied gas. The
spatial resolution of $5''$ is indeed crucial for resolving the physical
scale comparable to the Coulomb mean free path ($\sim 20$ kpc) of
electrons and protons in distant clusters. ALMA will also be able to
simultaneously identify and remove point sources that often contaminate
the diffuse SZE.

\section{Unresolved Structures of the Universe}
\label{sec-unres}

Recent developments of data sets over large sky areas have opened
several possibilities of probing yet unresolved structures of the
Universe by means of the SZE.

\begin{figure}[t]
\centering\includegraphics[width=76mm]{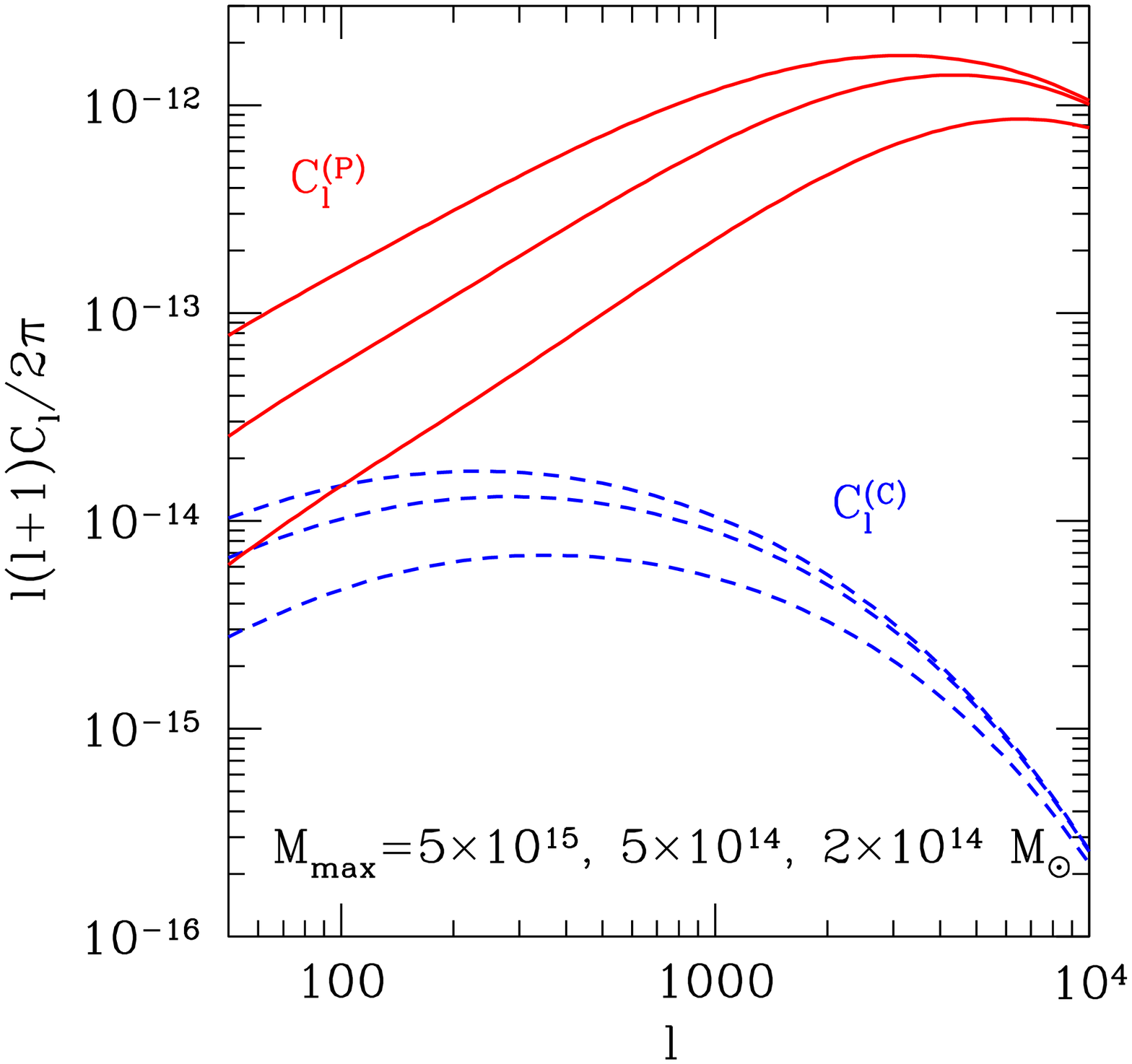}
\centering\includegraphics[width=76mm]{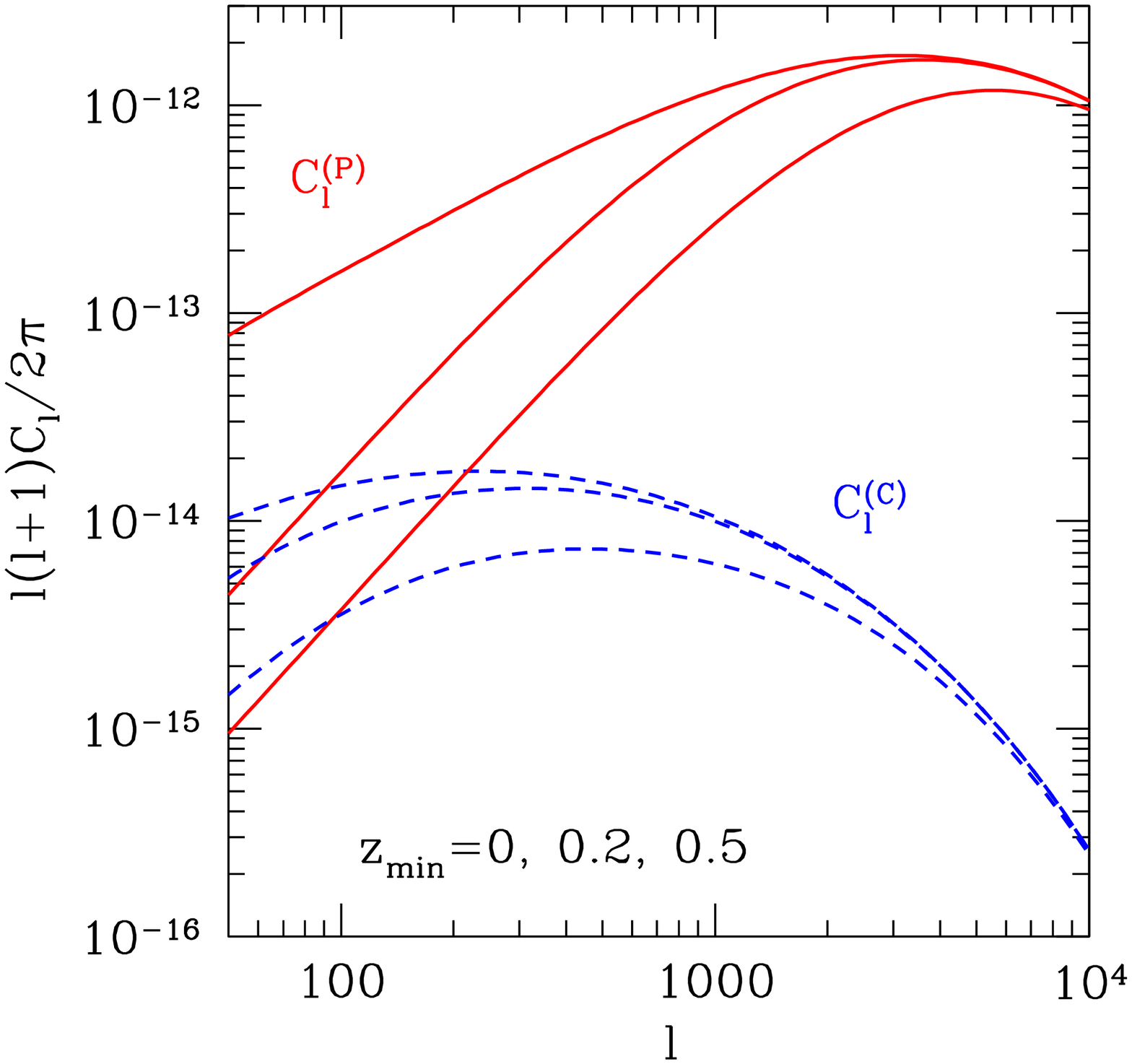}
\caption{Predicted angular power spectra of the Compton y-parameter from
the Poisson component (solid lines) and the correlation component
(dashed) in the conventional $\Lambda$CDM model. {\it Left}:
Contributions from clusters with masses lower than $M_{\rm max} = 5
\times 10^{15}$ (top), $5 \times 10^{14}$ (middle), and $2 \times
10^{14}$ (bottom) for $M_{\rm min}=10^{13} M_\odot$ and $z_{\rm
min}=0$. {\it Right}: Contributions from clusters at redshifts higher
than $z_{\rm min}=0$ (top), $0.2$ (middle), and $0.5$ (bottom) for
$M_{\rm min}=10^{13} M_\odot$ and $M_{\rm max}= 5 \times 10^{15}
M_\odot$. All the masses correspond to $M_{500}$.}  \label{fig-szpower}
\end{figure}

First, it has long been suggested that the integrated thermal SZE
signal, including low-mass clusters and groups of galaxies, contributes
to the CMB temperature anisotropies at sub-degree angular scales
\cite{ck88,ss88,markevitch92,makino93,bartlett94}.  The angular power
spectrum of the Compton y-parameter can be written as $C^{\rm yy}_l =
C_l^{\rm yy(P)} + C_l^{\rm yy(C)}$, where $C_l^{\rm yy(P)}$ is the
contribution from the Poisson noise and $C_l^{\rm yy(C)}$ is from
correlation among the sources. Employing the Limber's approximation
\cite{limber53}, one can write down these terms as \cite{kk99}
\begin{eqnarray}
  \label{eq:Cp}
  C_l^{\rm yy(P)} &=& \int_{z_{\rm min}}^{\infty} dz \frac{dV}{d\Omega dz}
                \int_{M_{\rm min}}^{M_{\rm max}} dM
                \frac{dn(M,z)}{dM} \left|\tilde{y}_l(M,z)\right|^2,\\
  \label{eq:Cc}
  C_l^{\rm yy(C)} &=& \int_{z_{\rm min}}^{\infty} dz \frac{dV}{d\Omega dz}
                       P_{\rm m}\left(k,z\right)
                       \left[\int_{M_{\rm min}}^{M_{\rm max}} dM
		       \frac{dn(M,z)}{dM}b(M,z) \tilde{y}_l(M,z)
		       \right]^2,
\end{eqnarray}
where $P_{\rm m}(k,z)$ is the 3D matter power spectrum, $k=l/
(1+z)/d_{\rm A}$ is the comoving wave number, $b(M,z)$ is the linear
bias factor of dark matter halos \cite{mw96,sheth99}.  The 2D angular
Fourier transform of the Compton y-parameter is given by \cite{ks02}
\begin{eqnarray}
 \tilde{y}_l (M,z) = 4\pi d_{\rm A}(z) \frac{\sigma_{\rm T}}{m_{\rm e}c^2} 
\int_0^{\infty} P_{\rm e}(\phi,M,z) 
\frac{\sin(l\phi)}{l\phi} \phi^2 d\phi,  
\end{eqnarray}
where $P_{\rm e}(\phi,M,z)$ is electron pressure at an angular radius
$\phi$ from the center of a cluster of mass $M$ at redshift $z$.  Figure
\ref{fig-szpower} shows an updated version of predictions by \cite{kk99}
in the conventional $\Lambda$CDM model using the mass function by
\cite{tinker08} and the pressure profile of equations (\ref{eq-gnfw})
and (\ref{eq-p500}) with the parameters given in \cite{arnaud10}. The
Poisson component is dominated by massive nearby clusters, which will be
identified individually. Once they are removed, the remaining power is
governed by low mass clusters at higher redshifts, with an increasing
contribution from the correlation component. The power at $l \nlt 1000$
will also provide a sensitive measure of $\sigma_8$, whereas it depends
on details of underlying pressure profile at higher multipoles (e.g.,
\cite{efstathiou12}).

The observed CMB power spectrum is dominated by primary anisotropies at
$l\nlt 2000$ and by radio sources or dusty star-forming galaxies at
higher multipoles (e.g., \cite{reichardt12,sievers13}).  Multi-frequency
observations are hence crucial for separating the SZE power from the
other components. Recent measurements of $C^{\rm yy}_l$ at $l \nlt 1000$
by Planck are in good agreement with the predictions similar to the one
mentioned above \cite{planck_y}.  Contribution of the kinematic SZE
is still uncertain and can arise from galaxy clusters
\cite{aghanim98,molnar00}, spatial variations of the ionized fraction
during cosmic reionization \cite{aghanim96,gruzinov98,knox98}, and
density fluctuations in the reionized universe (also called the
Ostriker-Vishniac effect) \cite{ostriker86,vishniac87,jaffe98}. The
latter two components potentially provide a unique probe of the cosmic
reionization history (e.g., \cite{zahn12}).

Second, luminous galaxies are often used to trace clusters or groups of
galaxies and their association with the thermal SZE have been studied by
stacking the data toward a large sample of bright galaxies
\cite{hand11,planck_group}. A clear correlation is found between the
stacked SZE flux and the stellar mass in the locally brightest galaxies
selected from the Sloan Digital Sky Survey down to $M_{\rm star} \sim
10^{11}M_\odot$, corresponding to the effective halo mass of $M_{500}
\sim 10^{13} M_\odot$ \cite{planck_group}. The gas content of such
low-mass halos is likely to account for a part of the missing baryon in
the Universe \cite{fukugita98}.

\begin{figure}[t]
\centering\includegraphics[width=80mm]{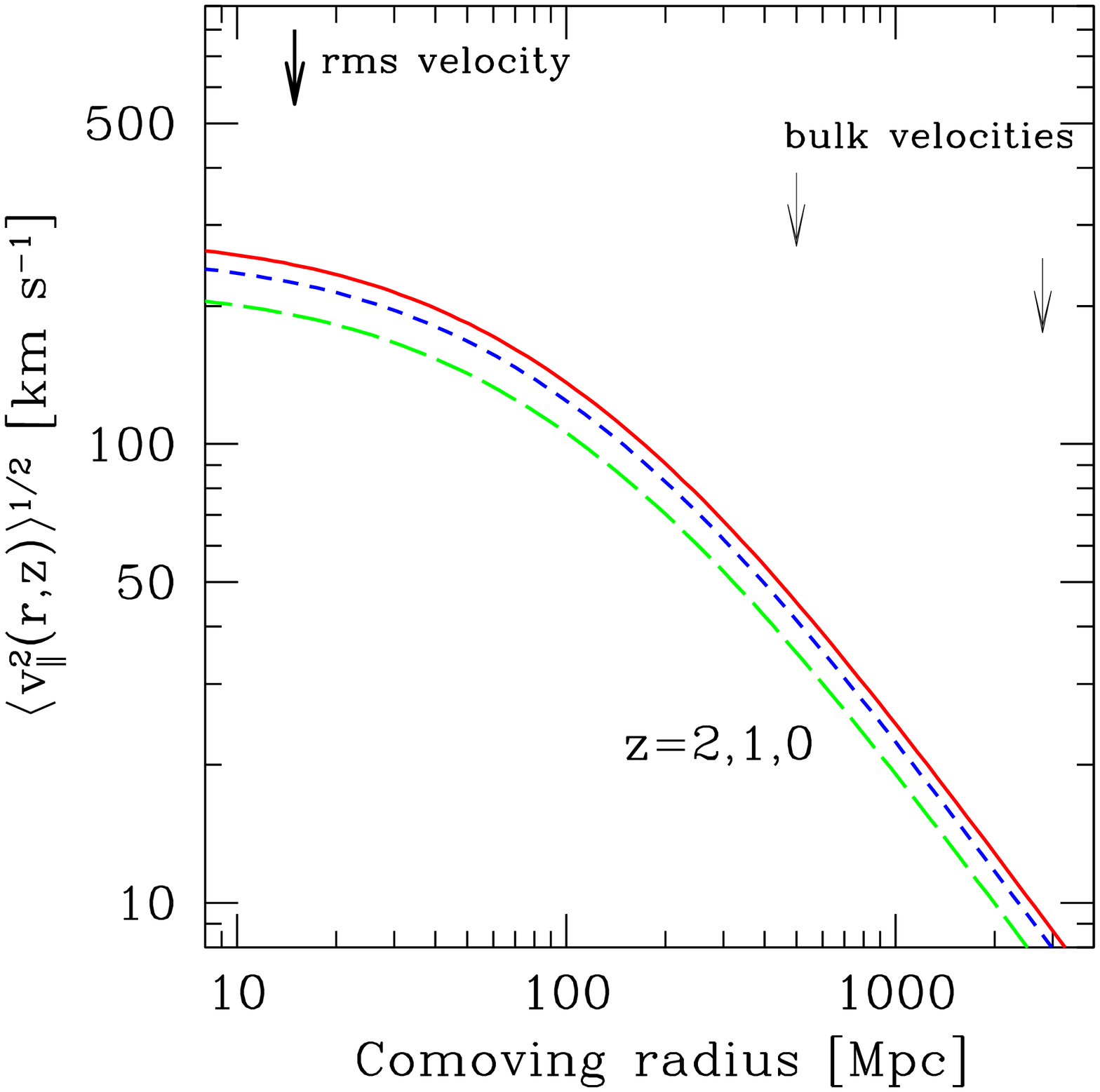} \caption{Linear
 theory predictions for the root-mean-square line-of-sight peculiar
 velocity over the comoving radius $r$ at $z=0$ (solid), $z=1$ (short
 dashed), and $z=2$ (long dashed) in the conventional $\Lambda$CDM
 universe. Thick and thin arrows mark the 2$\sigma$ upper limits on the
 rms velocity of massive clusters and the bulk velocities, respectively,
 measured by Planck \cite{planck_bulk}.}  \label{fig-pecvel}
\end{figure}

Finally, large-scale coherent motion of the matter can be studied by
means of the kinematic SZE. The linear perturbation theory predicts that
the variance of line-of-sight peculiar velocities induced by surrounding
density fluctuations is
\begin{eqnarray}
 \langle v_{\parallel}^2(r,z) \rangle = \frac{H_0^2 E(a)^2 a^2 }{6\pi^2}
\int_0^\infty \left(\frac{d\ln D}{d\ln a}\right)^2 
P_{\rm m}(k,z) |\tilde{W}_r(k) |^2 dk,
\label{eq-vpec}  
\end{eqnarray}
where $\tilde{W}_r(k)=3[\sin(kr)-kr\cos(kr)]/(kr)^3$ is the 3D Fourier
transform of the real-space top-hat filter over a comoving sphere of
radius $r$. Figure \ref{fig-pecvel} illustrates that the predicted
root-mean-square (rms) velocity is $200 \sim 300$ km\:s$^{-1}$ at $r=10$
Mpc corresponding to the enclosed mass of $M\simeq 2\times
10^{14}M_\odot$ and drops rapidly at larger radii with little redshift
dependence at $z<2$. Note that numerical simulations suggest that the
rms velocity of cluster-sized halos tends to be larger than the linear
theory prediction by $20 \sim 70 \%$ (e.g. \cite{hamana03}). While the
kinematic SZE of this amount of velocity is hard to measure for
individual clusters, a statistical detection of the mean pair-wise
velocity has been reported using the ACT 148 GHz data for a sample of
clusters and groups traced by 5000 luminous galaxies at $0.05 < z < 0.8$
\cite{hand12}. It has also been suggested that stacking the all-sky CMB
data toward known galaxy clusters will give a measure of the bulk flow
\cite{kashlinsky00}.  Recent Planck data place $2\sigma$ upper limits on
the rms radial velocity of $800$ km\:s$^{-1}$ for a sample of 100
massive clusters at $\langle z \rangle \sim 0.18$ and on the local bulk
flow velocity of $250$ km\:s$^{-1}$ within $\sim 3$ Gpc
\cite{planck_bulk} as marked in Figure \ref{fig-pecvel}. While the
limits are still weak, these measurements are consistent with
predictions in the $\Lambda$CDM universe.

\section{Summary}

Extensive efforts well over four decades have now established the SZE as
an indispensable tool in cosmology and astrophysics. Being one of the
major foregrounds of the CMB, the SZE not only plays a key role in
recovering correctly the primary anisotropies, but also offers unique
cosmological tests on its own. They include measurements of the
evolution of the CMB temperature, distances to high redshifts that are
entirely free from the cosmic distance ladder, the absolute numbers and
the power spectra of galaxy clusters, and large-scale motions of the
Universe. It should be noted that their accuracy critically depends on
our understanding of the physics of galaxy clusters and structure
formation, which the SZE observations have also been improving, e.g., by
finding high velocity cluster mergers, measuring pressure profiles, and
detecting the gas in low-mass halos. Perhaps the most noticeable
progress over the last decade or so is that the SZE measurements have
started to achieve their own discoveries independently of any other
means. This has made the SZE a truly complementary probe to X-ray
observations in the studies of cosmic plasma. A number of outcomes from
large area surveys and pointed observations by existing instruments are
also underway. It is highly anticipated that future SZE measurements
from both grounds and the space will continue to provide us new insights
into our Universe.

\section*{Acknowledgments}

We thank Issha Kayo, Eiichiro Komatsu, Yasushi Suto, and Keiichi Umetsu
for useful discussions and comments. We also thank the referees for
their careful reading of the manuscript and helpful suggestions. This
work is supported in part by the Grants-in-Aid for Scientific Research
by the Japan Society for the Promotion of Science (25400236).

\end{document}